\RequirePackage{etex}
\documentclass[aps,prb,twocolumn,showpacs,10pt]{revtex4-1}
\usepackage{amssymb}
\def\Title{From Disordered Quantum Walk to Physics of Off-diagonal Disorder}

\usepackage{graphicx}
\usepackage{amsmath,amsfonts,amssymb,dsfont,color,bbm,xcolor}
\usepackage{array}
\usepackage[thinlines]{easytable}
\usepackage{bbold}
\usepackage{natbib}
\usepackage{multirow}
\usepackage{booktabs}
\usepackage[colorlinks]{hyperref}
\usepackage[percent]{overpic}

\newcommand{\avg}[1]{\overline{#1}}
\newcommand{\expct}[1]{\left\langle#1\right\rangle}
\newcommand{\ket}[1]{\left|#1\right\rangle}
\newcommand{\bra}[1]{\left\langle #1 \right|}
\newcommand{\tmatrix}[2]{\begin{pmatrix} #1 \\ #2 \end{pmatrix}}

\newcommand{\be}{\begin{equation}}
\newcommand{\ee}{\end{equation}}

\newcommand{\bv}{\textbf{v}}

\newcommand{\bI}{\boldsymbol{I}}
\newcommand{\bP}{\boldsymbol{P}}
\newcommand{\SC}{\widetilde{C}}

\newcommand{\vt}{\vartheta}
\newcommand{\rmd}{\mathrm{d}}

\newcommand{\udt}{\text{DT}}
\newcommand{\tw}{\Delta}

\newcommand{\ip}[2]{\left\langle #1 | #2\right\rangle} 

\begin{document}

\bibliographystyle{apsrev4-1}

\author{Qifang Zhao}
\affiliation{Department of Physics and Centre for Computational Science and Engineering, National University of Singapore, Singapore 117546, Republic of Singapore}

\author{Jiangbin Gong}
\affiliation{Department of Physics and Centre for Computational Science and Engineering, National University of Singapore, Singapore 117546, Republic of Singapore}
\title{\Title}
\date{\today}

\begin{abstract}
Systems with purely off-diagonal disorder have peculiar features such as the localization-delocalization transition and long-range correlations in their wavefunctions. To motivate possible experimental studies of the physics of off-diagonal disorder,
we study in detail disordered discrete-time quantum walk in a finite chain, where the diagonal disorder can be set to zero by construction. Starting from a transfer matrix approach, we show, both theoretically and computationally, that
the dynamics of the quantum walk with disorder manifests all the main features of off-diagonal disorder.
We also propose how to prepare a remarkable delocalized zero-mode from a localized and easy-to-prepare initial state using an adiabatic protocol that increases the disorder strength slowly.  Numerical experiments are also performed with encouraging results.
\end{abstract}

\maketitle
\section{Introduction}
\label{sec:QW_intro}


Quantum walk (QW) has been a subject of great theoretical and experimental interests.
Among many QW protocols, discrete-time QW is the simplest~\cite{Aharonov1993}, where
it can be seen clearly how QW can differ strongly from classical random walk due to
quantum interference effects.
For example, an initially localized state in QW will spread ballistically,
which is much faster than classical random walk whose mean square displacement
is proportional to time.  Due to this feature, one potential application of QW models
 is towards a fast search algorithm ~\cite{Kempe2003} in quantum computation~\cite{Farhi1998}.
As a very recent direction, QW is shown to be useful
in understanding topological phases of matter in periodically driven systems~\cite{Kitagawa2010,Ho2012}.

On the experimental side, two early QW experiments
in 2005 used either linear optical elements~\cite{Do2005} or nuclear-magnetic
resonance systems~\cite{Ryan2005}. Since 2007, a variety of physical systems has been exploited
to realize QW, including trapped ions~\cite{Schmitz2009,Zahringer2010},
trapped atoms in a spin-dependent optical lattice~\cite{Karski2009},
photons in an optical waveguide array~\cite{Perets2008,Peruzzo2010,Owens2011,Sansoni2012},
and photonic walks with interferometers~\cite{Zhang2007,Broome2010,Schreiber2010}.
Very recently, a photonic quantum walk without interferometers was realized~\cite{Cardano2014},
in which photons walk in the orbital angular momentum space.

The topic of this work is on QW in the presence of some disorder.
Previously, it was numerically found that some behavior of disordered QW seems to reflect
the physics of off-diagonal disorder (ODD)\cite{Obuse2011} in condensed-matter physics. The so-called
ODD was first
noticed in studies of one-dimensional (1D) tight-binding models (TBMs) with
random hopping potential and constant on-site potential~\cite{Theodorou1976,Eggarter1978}.
Compared with the more familiar disorder model where the on-site potential (diagonal term in the lattice-site representation)
is random but the hopping is constant, ODD leads to peculiar physics,
such as delocalization at zero energy, power-law wavefunction correlation, and so
on~\cite{Dyson1953,Theodorou1976,Eggarter1978,Zirnbauer1994,Lee1994,Kondev1997,Balents1997,Shelton1998,Unanyan2010,Edmonds2012}.
Specifically, the localization length $\ell(\omega)$ in 1D TBM with pure ODD
 is related to energy $\omega$ via
\begin{equation}
    \label{Off_diagonalLen}
   \ell(\omega) \propto |\ln\omega|.
\end{equation}
As the energy $\omega$ approaches $0$, the localization length $\ell$ diverges,
indicating a delocalization transition at $\omega=0$.
At the same time, singularity in the density of states (DOS) emerges at $\omega=0$, with the explicit DOS expression given by
\begin{equation}
     \label{Off_diagonalDOS}
     \rho(\omega) \propto |\omega \ln^3 \omega|^{-1}.
\end{equation}
Furthermore, the delocalized eigenstate has an unusual long-range correlation.
It is shown that its ensemble averaged two-point correlation decays polynomially with the
exponent $-3/2$ under the condition of strong disorder and large two-point
separation~\cite{Balents1997,Steiner1998,Shelton1998}. It was pointed out earlier
that this is a manifestation of the actual stretched exponential-decay
profile of the wave function~\cite{Fleishman1977,Soukoulis1981,Markos1988,Bovier1989}, i.e., $\psi(x) \propto \exp(-\widetilde{\gamma} |x-x_0|^{1/2})$, where	$\widetilde{\gamma}$ is a constant.
One may naively say that a wavefunction like this is quite localized. However,
its Lyapunov exponent is apparently zero (which indicates that the state is delocalized~\cite{Fleishman1977}) because there is no exponential localization behavior.

As we have learnt from decades of studies, quite a few theoretical models with disorder can be used to
manifest and digest the physics of ODD.  Such models include a special
disordered linear chain of harmonic oscillators investigated by
Dyson~\cite{Dyson1953,Lieb1961,Smith1970}, a 1D Dirac model with random
mass and some types of disordered 1D spin chains~\cite{Balents1997,Steiner1998,Shelton1998},
2D Dirac fermions subject to a random vector potential~\cite{Ludwig1994},
a 1D random hopping model consisting of several parallel bipartite
sublattices~\cite{Brouwer2002}, systems with correlated off-diagonal
disorder~\cite{deMoura1999,Cheraghchi2005} or random long-range hopping~\cite{Zhou2003},
and graphene with ODD~\cite{Zeuner2013}.
In contrast to these theoretical developments,  experimental progresses on the physics of ODD
have been rather limited. Doped $\text{CuGeO}_3$ is effectively
a disordered spin-Peierls system possessing ODD~\cite{Hase1993,Oseroff1995,Hase1996,Masuda1998,Wang1999,Nakao1999}.
There phenomena like phase transitions and
long-range orderings were believed to be related to the physics of ODD. However,
direct observation of physical properties like the correlation exponent $-3/2$ was not possible in such a system.
Other than spin-chain realizations, few experiments concerning ODD
were reported. We note a possible experimental approach based on cold atoms
under the so-called tripod scheme~\cite{Unanyan2010,Edmonds2012}, but the actual experiment has not been done.
Only very recently, Keil {\it et al} demonstrated that a chain
of optical waveguides could be used to realize an effective 1D Dirac model with random mass~\cite{Keil2013}.
In particular,  with coupled series of optical chains, the authors of Ref.~\cite{Keil2013} observed the long
range correlation (in a certain range) characterized by the correlation exponent $-3/2$.


To motivate more possible experimental studies of ODD models and to demonstrate one more
 promising application of QW, we consider in this work a discrete-time QW in a finite chain
(for simplicity we refer to it as ``QW'' throughout the paper) and reveal theoretically
how this problem is closely connected with the issue of ODD.
Our work is inspired by an early numerical study by Obuse and Kawakami~\cite{Obuse2011},
which showed clear signatures of the physics of ODD in disordered QW.
Specifically, we first analytically demonstrate the explicit connection
between a TBM with ODD and disordered QW.
In so doing we focus on a specific delocalization transition energy, the
zero quasi-energy, which was also considered in Ref.~\cite{Obuse2011}.
We then show how some simple adiabatic protocols, starting from
an exponentially localized 0-mode (i.e., the 0 quasi-energy eigenstate),
can be converted to a peculiar 0-mode possessing the physics of ODD, with
satisfactory fidelity and relatively short duration of the protocol.
As such, we may make use of some existing QW experimental set-ups to observe
the unique physics of ODD. Indeed, our numerical experiments
indicate that the results agree with theoretical predictions very well, including
the $-3/2$ correlation exponent.
One advantage of this QW approach is that the diagonal disorder
does not exist by construction, so that the results are free of
 any possible contamination due to diagonal disorder.

This paper is organized as follows. In Sec.~\ref{sec:QW_setup},
we will introduce a model of disordered QW in a finite chain.  Analysis of the model
is based on the transfer
matrix formalism. Sec.~\ref{sec:EDOS} is devoted to some formal connections
between our QW model and a TBM with ODD.
In Sec.~\ref{sec:QW_expt} we shall focus on the preparation of special states
that best manifest the peculiarities of ODD. The associated results from our
numerical experiments will be also presented and discussed. Sec.~\ref{sec:QW_con}
concludes this work.
\section{Disordered QW in a Finite Chain}
\label{sec:QW_setup}
The standard discrete-time QW is defined via a
single particle with two internal degrees of freedom.
For convenience, we refer to its internal states as ``spin-up'' and ``spin-down''.
The QW protocol consists of two operations,
a rotation of spin through operator $R$, followed by a shift operation by $S$.
Without loss of generality, we consider a rotation around $y$ axis
by an angle $2\theta$, such that $R=e^{-i\theta\sigma_y}$:
\begin{equation}
    \label{QW_rot_angle}
			R(\theta)= \begin{pmatrix}
        \cos\theta   &  -\sin\theta \\
	     \sin\theta   &  \cos\theta
                 \end{pmatrix}.
\end{equation}
The operator $R$ rotates the spin at each site, and then the spin-up
component walks to the right, whereas the spin-down component walks
the left.  Such spin-dependent shift operation is implemented via the operator $S$:
\begin{equation}
   S=\sum^{\infty}_{n=-\infty}{\left(\ket{n+1}\bra{n}\otimes\ket{\uparrow}\bra{\uparrow}+\ket{n-1}\bra{n}\otimes\ket{\downarrow}\bra{\downarrow}\right).}
\end{equation}
The overall  one-step quantum walk operator (without disorder) is then given by
\begin{equation}
     \label{Floquet_DTQW}
     U_\udt\equiv S\left(\sum_n{\ket{n}\bra{n}} \otimes R\right).
\end{equation}

The above described QW can be restricted
to a finite regime~\cite{Kitagawa2012,Obuse2011,Asboth2012} through
total-reflection coin operators $R_{\pm}$ at two boundaries, with $R_{\pm}$ defined as
\begin{equation}
    R_{\pm} = \begin{pmatrix}
		               0   &  \mp 1 \\
							 \pm 1 &  0
		          \end{pmatrix}
						= \begin{pmatrix}
		               \cos(\pm \frac{\pi}{2})   &  -\sin(\pm \frac{\pi}{2}) \\
							     \sin(\pm \frac{\pi}{2})   &  \cos(\pm \frac{\pi}{2})
		          \end{pmatrix}.
\end{equation}
Note that $R_{\pm}$ preserves the particle-hole symmetry and conserves the probability inside a finite
QW chain. 
$R_{\pm}$ turns spin-down to spin-up, and vice versa.
Since the coin operators at two boundaries can be either $R_+$ or $R_-$, we could
have 4 choices of boundaries as $[R(\theta_0), R(\theta_{N+1})]=(R_{\pm}, R_{\pm})$.
In the following we mainly choose $(R_-, R_+)$ as our boundary condition.
Studies of other boundary conditions can be found in Appendix~\ref{sec:MB}.
As depicted in Fig.~\ref{fig:QW_setup}, our QW model has totally $N+2$ sites,
with $N$ of them being bulk sites.

Next we introduce disorder to the QW model,  by considering
a perturbation to the local rotation angles $\theta_n$, i.e.,
\begin{equation}
     \label{QW_BulkAngle}
      \theta_n=\tilde{\theta}+\delta_n \text{  for  } n=1,\ 2,\ ...\ N.
\end{equation}
Here $\tilde{\theta}$ is identical for different sites $n$, while
$\delta_n\in[-\tw,\tw]$ may differ from site to site, giving rise to a disordered QW on a finite number
of sites.

\begin{figure}
\centering
   \includegraphics[width=\linewidth]{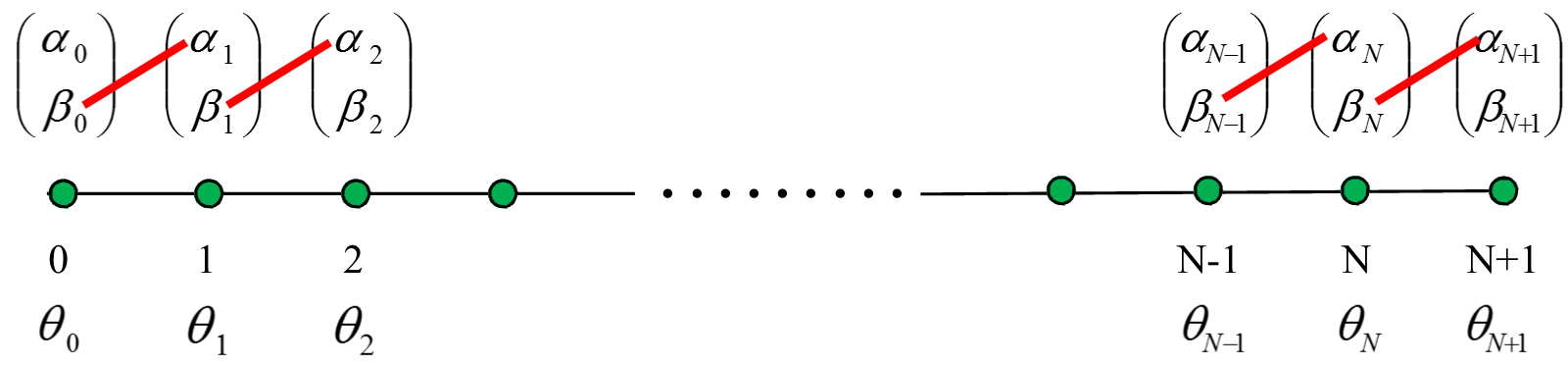}
   \caption[Set-up of quantum walk in a finite chain]
	  {(Color online) Set-up of our finite-chain QW with disorder, with totally $N+2$ sites,
	where site $0$ and $N+1$ are the boundary sites with reflection operators $R_-$
	and $R_+$. Rotation operators of bulk sites with $n=1,2\ldots N-1,N$ depend on the local
	angle $\theta_n$, which fluctuate from site to site. The red slashes connect
	spin components $\beta_n$ and $\alpha_{n+1}$, as they form the new ``spinor''
	in our transfer matrix formalism elaborated in our main text.
		}
   \label{fig:QW_setup}
\end{figure}

For such a finite-site QW system with a disordered bulk
specified by $\theta_n$, we can still define a mapping operator $U$, which can be adpated
from the $U_\udt$ in Eq.~\eqref{Floquet_DTQW} [that is, $R(\theta)\rightarrow \prod_{n} R(\theta_n)$].
In representation of different QW sites, $U$ can be expressed explicitly as a $2(N+2)\times 2(N+2)$ matrix.
As a mapping operator, $U$ is unitary with eigenvalue $e^{i\omega}$:
\begin{equation}
     \label{Sch_eq}
     U\ket{\psi}=e^{i\omega}\ket{\psi},
\end{equation}
where $\omega$ is the quasi-energy eigenvalue of $U$, $\ket{\psi}$ is the associated eigenstate characterized by
\begin{equation}
     \label{explitPsi}
     \ket{\psi}=(\alpha_0\ \beta_0\ \alpha_1\ \ldots\ \alpha_{N+1}\ \beta_{N+1})^\text{T},
\end{equation}
with $(\cdots)^\text{T}$ being the transpose operation.
Because of the special choices of rotation operators at two boundaries,
the first and last rows, and the first and last columns of
$U$ have entries 0 only.  Upon removing these rows and columns,
$U$ becomes a $2(N+1)\times 2(N+1)$ matrix. Correspondingly,
the entries $\alpha_0$ and $\beta_{N+1}$ in the eigenstate $\ket{\psi}$
can be also removed.

\subsection{Transfer matrix formalism}
\label{sec:QW_solving}
In solving Eq.~\eqref{Sch_eq}, one obtains the following recursive relation
between the entries of the eigenstate $\ket{\psi}$:
\begin{equation}
     \label{Chain_rela}
     \begin{cases}
		       \alpha_n e^{i\omega} & = \alpha_{n-1} \cos\theta_{n-1} - \beta_{n-1} \sin\theta_{n-1},\\
					 \beta_n e^{i\omega} & = \alpha_{n+1} \sin\theta_{n+1} + \beta_{n+1} \cos\theta_{n+1},
		 \end{cases}
\end{equation}
with $n\in[1,N]$. Such relations can be expressed in the following
matrix form:
\begin{equation}
    \label{TMM_U}
    \tmatrix{\beta_n}{\alpha_{n+1}}
		  =T_n \tmatrix{\beta_{n-1}}{\alpha_n},
\end{equation}
with
\begin{equation}
    \label{TMM_UII}
    T_n = \begin{pmatrix}
		               e^{i\omega} \sec\theta_n   &  -\tan\theta_n \\
							     -\tan\theta_n &  e^{-i\omega} \sec\theta_n
		    \end{pmatrix}.
\end{equation}
Here $T_n$ is the transfer matrix~\cite{Obuse2011} at site $n$.
In Eq.~\eqref{TMM_U}, the neighboring spinors' components
$\beta_{n-1}$ and $\alpha_n$ form the new ``spinors'' (See Fig.~\ref{fig:QW_setup}), and
they are chained through local transfer matrices. Disordered parameter $\theta_n$
and quasi-energy $\omega$ are contained in these matrices.
This allows us to deal with disorder explicitly. This is one known advantage
of the transfer matrix formalism (TMF)~\cite{Markos2008,Muller2011}.

Given the chain relation between entries of the eigenstate $\ket{\psi}$ in
Eq.~\eqref{TMM_U}, we still need to handle the boundary situations with care, i.e.,
$\tmatrix{\beta_0}{\alpha_1}$ and $\tmatrix{\beta_{N}}{\alpha_{N+1}}$.
By setting $n$ in Eq.~\eqref{Chain_rela} to be $0$ and $N$, we obtain
\begin{equation}
     \label{EigenState_boudary}
     \begin{cases}
		       \alpha_1 e^{i\omega} & = \alpha_{0} \cos\theta_{0} - \beta_{0} \sin\theta_{0},\\
					 \beta_N e^{i\omega} & = \alpha_{N+1} \sin\theta_{N+1} + \beta_{N+1} \cos\theta_{N+1},
		 \end{cases}
\end{equation}
which further reduce to
\begin{equation}
    \label{DTQW_spinor}
    \tmatrix{\beta_0}{\alpha_1} = c_0 \tmatrix{e^{i\omega}}{-\sin\theta_0}, \quad \tmatrix{\beta_{N}}{\alpha_{N+1}} = c_{N} \tmatrix{\sin\theta_{N+1}}{e^{i\omega}}.
\end{equation}
Using the boundary conditions in Eq.~\eqref{DTQW_spinor}
, the chain relation in Eq.~\eqref{TMM_U}, as well as
$\theta_0=-\pi/2$ and $\theta_{N+1}=\pi/2$, we finally
obtain the following equation that carries all the information of
Eq.~\eqref{Sch_eq}:
\begin{equation}
\label{RecurRela}
      c_{N} \tmatrix{1}{e^{i\omega}} = T_{N} \cdot T_{N-1} \cdot\cdots T_2\cdot T_1 \cdot c_0 \tmatrix{e^{i\omega}}{1}.
\end{equation}
For a specific realization of disorder, only
particular values of the quasi-energy $\omega$ satisfy Eq.~\eqref{RecurRela}.
 The coefficients $c_{N}$ and
$c_0$ can be determined from Eq.~\eqref{RecurRela} and the normalization
of $\ket{\psi}$.

To conclude, the TMF reduces a matrix equation with
dimension $2(N+1)\times 2(N+1)$ [Eq.~\eqref{Sch_eq}] to a chained matrix equation connecting
$N$ matrices, each of dimension $2\times 2$ [Eq.~\eqref{RecurRela}].
This framework will be used later.
Indeed, in the following we will not return to the original Eq.~\eqref{Sch_eq} but
just focus on Eq.~\eqref{RecurRela}.

\subsection{Special quasi-energies and the implication of ODD}
\label{sec:zeroE}
By observing the transfer matrix in Eq.~\eqref{TMM_UII}, we notice that $\omega=0,\pm\pi/2,\pi$
are special quasi-energies. For example, when $\omega=0$, the transfer matrix reduces to
\begin{equation}
  \label{QW_Tn_0e}
  T_n=\sec\theta_n\cdot \bI -\tan\theta_n \sigma_x,
\end{equation}
where $\bI$ is the identity $2\times 2$ matrix.
Such simple transfer matrices can be exactly diagonalized in the basis of
$\sigma_x$, so that the product of all the transfer matrices can be easily calculated.
This being the case, whether $\omega=0$,$\pm\pi/2$, or $\pi$ satisfies Eq.~\eqref{RecurRela} can be checked without difficulty.
If $\omega$ is not equal to one of these special values, then it is virtually impossible to
analytically check Eq.~\eqref{RecurRela} because the product of these transfer matrices is hard to evaluate.

If $\omega$ assumes one of these special values, the corresponding eigenstates can be also analyzed in a straightforward manner.
Take again the case of $\omega=0$ as an example. When $\omega=0$, from Eq.~\eqref{QW_Tn_0e} we get
\begin{equation}
   \label{QW_TMM_diag}
   \begin{split}
      \prod^{N}_{n=1} T_n & = \frac{1}{2}\left(\lambda_+ + \lambda_-\right)\bI + \frac{1}{2}\left(\lambda_+ - \lambda_-\right)\sigma_x,\\
			\text{ with } \lambda_+ & = \lambda^{-1}_- = \prod^{N}_{n=1} \tan\left( \frac{\pi}{4} - \frac{\theta_n}{2}\right).
	 \end{split}
\end{equation}
And the ``spinors'' at both ends of $\ket{\psi}$ are proportional to $\tmatrix{1}{1}$,
i.e., the eigenvector of $\sigma_x$, obtained from Eq.~\eqref{DTQW_spinor}.
Substituting Eq.~\eqref{QW_TMM_diag} into Eq.~\eqref{RecurRela}, we get
\begin{equation}
     \label{QW_TMM_diagII}
      \tmatrix{1}{1} = \frac{c_0}{c_N} \lambda_+ \tmatrix{1}{1},
\end{equation}
which obviously holds by an appropriate choice of $c_0/c_N$.
Therefore, $\omega=0$
is indeed a quasi-energy solution of the disordered QW system.

In Eq.~\eqref{QW_TMM_diag}, if $\theta_n$ fluctuates around 0
or $\pi$ (i.e., $\tilde{\theta}=0$ or $\pi$),
$\ln |\lambda_+|$ will follow unbiased diffusion process around 0, so
$|\lambda_+|\approx 1$ for large $N$, which means that exponential decay
of the eigenstate $\ket{\psi}$ does not occur.
This quantitative analysis resembles that of off-diagonal
disordered TBM~\cite{Theodorou1976,Eggarter1978}, so we suspect that
our model also displays the physics of ODD. Indeed,
later in Sec.~\ref{sec:EDOS} we shall show that
$\omega=0$ is the localization-delocalization transition quasi-energy,
and Dyson's singularity emerges there, provided that
$\theta_n$ takes values randomly from a box distribution
$[-\tw,\tw]$. If $\theta_n$ fluctuates around values other than 0 or $\pi$, $|\lambda_+|$ will increase
or decrease exponentially, resulting in the localized 0- or $\pi$-mode, which
we believe, is related to those topologically protected edge states currently
being studied~\cite{Kitagawa2012}.

In the rest of this paper, we focus on the quasi-energy $\omega=0$ and
quasi-energies in its vicinity. In Appendix~\ref{sec:QW_other_quasi},
we shall discuss those cases with quasi-energy values other than 0 or $\pi$.

\section{Physics of ODD}
\label{sec:EDOS}
As introduced in Sec.~\ref{sec:QW_intro}, ODD is quite
different from diagonal disorder and leads to peculiar properties.
For our QW model, here we attempt to derive its DOS and localization length, keeping mind that
it is possible for a delocalization transition to occur at some special quasi-energy values.

\subsection{Analyzing quasi-energy values}
\label{sec:DOS}
We start with Eq.~\eqref{RecurRela} by considering its alternative form after some transformations:
\begin{equation}
\label{RecurRelaBN3}
\begin{split}
		 \tmatrix{1}{0}
       & = c
		        \begin{pmatrix}
		               \cos \omega   &  i\sin \omega \\
							    i\sin \omega   &  \cos \omega
		        \end{pmatrix}				
			 \cdot \bP \cdot \tmatrix{1}{0}, \text{ with } \\
		\bP
		& = \prod^{N}_{n=1} \left[
			      \begin{pmatrix}
		               \tan \vt_n   &  0 \\
							     0            &  \cot \vt_n
		        \end{pmatrix}
				\begin{pmatrix}
		               \cos \omega   &  i\sin \omega \\
							    i\sin \omega   &  \cos \omega
		    \end{pmatrix} \right],
\end{split}
\end{equation}
where $\vt_n = \frac{\pi}{4} - \frac{\theta_n}{2}$.
The detailed derivation can be found in Appendix~\ref{sec:EqTrans}.
Note that if and only if $\omega$ takes the actual quasi-energy value, then Eq.~\eqref{RecurRelaBN3} will be satisfied.
In particular,  it is now obvious to observe from Eq.~(\ref{RecurRelaBN3}) that $\omega=0$ is one quasi-energy value.
To derive DOS, we need to analyze other quasi-energy values allowed by Eq.~\eqref{RecurRelaBN3}.
 To that end we first re-interpret Eq.~\eqref{RecurRelaBN3}, which
is inspired by Schmidt's work~\cite{Schmidt1956} that treats spinors linked
by transfer matrices as vectors in a plane.

\begin{figure}[!ht]
\begin{center}$
\begin{array}{c}
\begin{overpic}[width=0.71\linewidth]{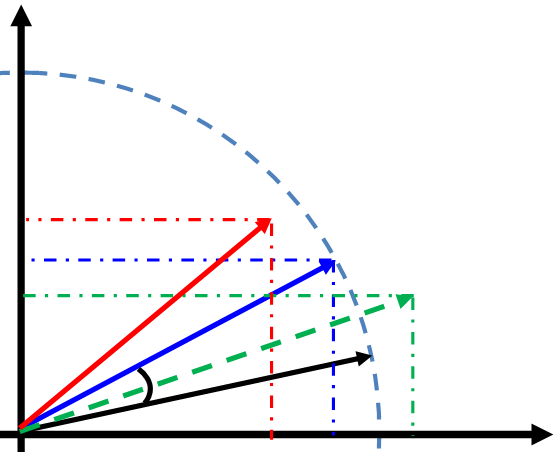} 
\put (80,70) {\Large$(a)$}
\put (7,75) {\Large $y$}
\put (90,6) {\Large $x$}
\put (-14,40) {\large $1.25y$}
\put (-8,33) {\large $y$}
\put (-12,25) {\large $0.8y$}
\put (41,-4) {\large $0.8x$}
\put (58,-4) {\large $x$}
\put (67,-4) {\large $1.25x$}
\put (28,11) {\huge $\omega$}
\put (41,44) {\LARGE $\bv_{n+1}$}
\put (59,33) {\LARGE $\bv_{\text{mid}}$}
\put (75,25) {\LARGE $\bv'_{n+1}$}
\put (66,15) {\LARGE $\bv_{n}$}
\end{overpic}
\\
\\
\begin{overpic}[width=0.7\linewidth]{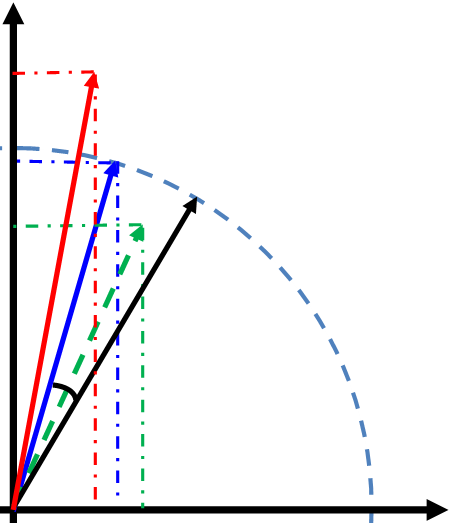}
\put (70,70) {\Large$(b)$}
\put (7,95) {\Large $y$}
\put (80,6) {\Large $x$}
\put (-14,84) {\large $1.25y$}
\put (-8,68) {\large $y$}
\put (-12,55) {\large $0.8y$}
\put (9,-4) {\large $0.8x$}
\put (21,-4) {\large $x$}
\put (25,-4) {\large $1.25x$}
\put (11,27) {\huge $\omega$}
\put (12,89) {\LARGE $\bv_{n+1}$}
\put (19,71) {\LARGE $\bv_{\text{mid}}$}
\put (23,57) {\LARGE $\bv'_{n+1}$}
\put (35,63) {\LARGE $\bv_{n}$}
\end{overpic} \\
\end{array}$
\end{center}
\caption[Re-interpretation of transfer matrices chain.]{
    (Color online) The operations in Eq.~\eqref{RecurRelaBN3} illustrated via a  complex plane with the$x$-axis denoting the real part (the first component of the spinor) and the $y$-axis denoting the imaginary part (the second component of the spinor).
    In the first quadrant, from top
		to bottom, the four vectors are $\bv_{n+1}$,
		$\bv_{\text{mid}}$, $\bv'_{n+1}$and
		$\bv_{n}$. $\SC_n\cdot \widetilde{R}$ acts on $\bv_n$ to get $\bv_{n+1}$ (if contracted)
		or $\bv'_{n+1}$ (if stretched). Specifically, $\widetilde{R}$ rotates $\bv_{n}$ by angle $\omega$
		to get $\bv_{\text{mid}}$; then $\SC_n$ will stretch or contract $\bv_{\text{mid}}$,
		In panel $(a)$, $\bv_{\text{mid}}$'s
		angle is less than $\pi/4$, while in panel $(b)$ its angle is larger than $\pi/4$.
		Hence the length of $\bv_{n+1}$ in panel $(a)$ is smaller than in panel $(b)$,
		whereas the opposite is true for $\bv'_{n+1}$.}
\label{fig:QW_VectorsInterpretation}
\end{figure}
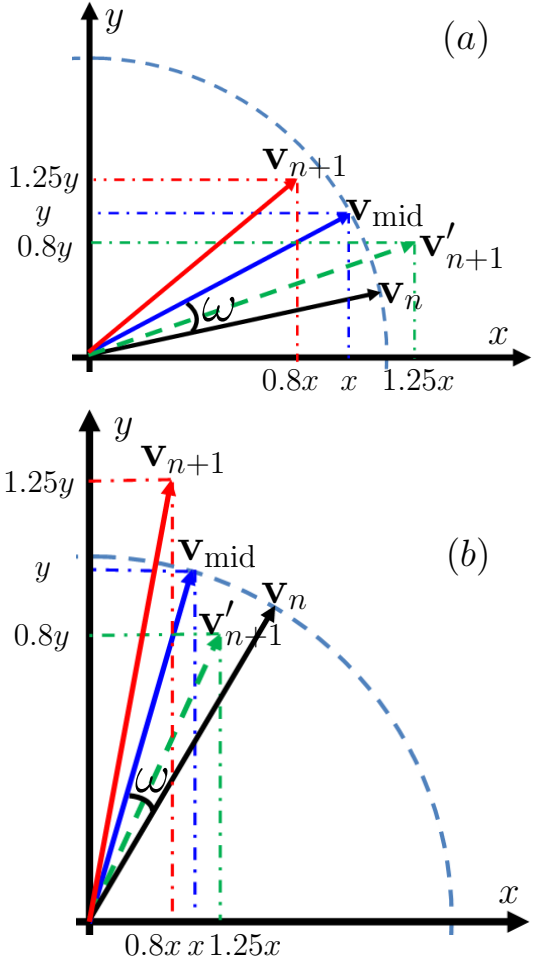

Let us consider a complex plane with $x$-axis denoting the real part,
while $y$-axis denoting the imaginary part. In Eq.~\eqref{RecurRelaBN3},
the initial ``spinor'' $\tmatrix{1}{0}$ can be treated as a vector lying in the
real axis with length 1 pointing in the positive direction. So from now on,
we refer to the ``spinor'' as a ``vector''. Let
\begin{equation}
    \label{QW_newTMF}
      \widetilde{R}= \begin{pmatrix}
		               \cos \omega   &  i\sin \omega \\
							    i\sin \omega   &  \cos \omega
		        \end{pmatrix}	
			\text{ and }
			\SC_n = \begin{pmatrix}
		               \tan \vt_n   &  0 \\
							     0            &  \cot \vt_n
		        \end{pmatrix},
\end{equation}
so $\widetilde{R}$ and $\SC_n$ do the  job 
of $\bP$ in Eq.~\eqref{RecurRelaBN3}.
Consider a vector $\bv_n = \tmatrix{x_n}{iy_n}$. Its angle with
respect to positive $x$-axis is $\phi_n$, and $\tan\phi_n = y_n/x_n$.
According to Eq.~\eqref{RecurRelaBN3}, we define
\begin{equation}
    \label{QW_newTMFII}
     \bv_{n+1} = \SC_n\cdot \widetilde{R} \cdot \bv_{n},
\end{equation}
with $n=1,\ 2,\ \cdots \ N$, and $\bv_1 = \tmatrix{1}{0}$. Hence, we can
interpret Eq.~\eqref{QW_newTMFII} (and Eq.~\eqref{RecurRelaBN3} thereafter)
as the following (see also Fig.~\ref{fig:QW_VectorsInterpretation}):
$\widetilde{R}$ rotates vector $\bv_n$ counter-clockwise by an angle $\omega$, followed
by stretching in $x$-coordinate by a factor $\tan\vt_n$ and $y$-coordinate by the factor
$\cot\vt_n$ (due to $\SC_n$), and then $\bv_{n+1}$ is reached with the following relation
     \begin{equation}
				      \label{RelaTan}
				      \tan\phi_{n+1} = \tan\left(\phi_n + \omega \right) \cot^2 \vt_n.
			\end{equation}

In Eq.~\eqref{RecurRelaBN3}, the initial vector $\bv_i$ and
final vector $\bv_f$ are both $\tmatrix{1}{0}$, and $\bv_i=\bv_1$,
$\bv_f=\widetilde{R}\cdot \bv_{N+1}$, so $\tan\phi_1 = \tan(\phi_{N+1}+\omega) = 0$.
As such, Eq.~\eqref{RecurRelaBN3} presents such a physical picture:
a vector initially located in positive $x$-axis is rotated and stretched or contracted, repeatedly,
and after a final rotation, it lands back on the $x$-axis. Therefore,
\begin{equation}
       \label{phiN}
         \phi_{N+1}+\omega=j\pi.
\end{equation}
Note that $\omega$ has
the period of $2\pi$, so we assume $\omega \in[-\pi,\pi]$.
Through interpreting Eq.~\eqref{RecurRelaBN3} this way,
we are now ready to derive the DOS near $\omega=0$.
Without loss of generality, we consider a small positive quasi-energy $\omega$.

Regarding the rotating and stretching and contracting processes, there are two
important factors to be noted. First, $\phi_n$ does not increase monotonically with respect to $n$.
$\phi_{n+1}$ could be smaller than $\phi_n$ (see Fig.~\ref{fig:QW_VectorsInterpretation}). However, $\phi_n$ has a
tendency to increase because the positive $\omega$ forces $\bv_n$ to
rotate counterclock-wise. Besides, a vector $\bv_n$
can never cross $x$ and $y$-axis clockwise. For example, if $\bv_n$
is inside the first quadrant, then $\tan\phi_n$ and $\cot^2\vt_n$
are positive, so  for $\tan\phi_{n+1}$ in Eq.~\eqref{RelaTan} to be negative
(i.e., crossing the axis), $\tan(\phi_n+\omega)$ must be negative.
Therefore, only the rotation $\widetilde{R}$ can bring a vector from one quadrant
to another, while the stretching and contracting operation $\SC_n$ cannot.
The vector $\bv_n$ can only drift away by crossing the positive $y$-axis.
Thus,  in Eq.~\eqref{phiN}, $j$ is always a positive integer.
Second, in a single realization of disorder, the following equation holds
\begin{equation}
    \label{QW_2eneries}
      \phi_{N+1}(\omega_b) > \phi_{N+1}(\omega_a) \text{  for  } \omega_b > \omega_a.
\end{equation}
To prove this relation, we show that given $\phi_n\geq\phi'_n$ and
$\omega>\omega'$, then $\phi_{n+1}>\phi'_{n+1}$. We assume that $\phi_{n}$ and $\phi'_{n}$
are quite close and and within the same quadrant, say the first quadrant. Then it is
easy to see that
\begin{equation}
    \begin{split}
      & \tan\phi_{n+1}-\tan\phi'_{n+1} \\
			=& \left[\tan(\phi_n + \omega)-\tan(\phi'_n+\omega')\right]\tan^2\vt_n>0,
		\end{split}
\end{equation}
so we get $\phi_{n+1}>\phi'_{n+1}$. This conclusion can be easily proved
in other quadrants, too.
Hence, starting with the same initial condition $\phi_1=0$ and same realization
of disorder, after
$N$ cycles, the associated $\phi_{N+1}(\omega)$ is a monotonous function of $\omega$.
This feature is checked in our numerical studies.

Given the two factors above, we can now count the number of states between quasi-energies $0$
and $\omega$. Suppose that the corresponding vector of $\omega$ sweeps an angle in-between
$j\pi$ and $(j+1)\pi$, then there exists $j$ quasi-energies $\omega_1\cdots\omega_j$ that
are the solution of the systems, and their vectors sweep angles $\pi\cdots j\pi$ correspondingly.
Therefore, the number of states between $0$ and $\omega$ is $j$, and specifically,
\begin{equation}
      \text{If } j\leq \frac{\phi_N(\omega) + \omega}{\pi} < j + 1,
\end{equation}
and
\begin{equation}
    \begin{split}
      & \omega_1 <\omega_2 <\cdots<\omega_k\cdots<\omega_{j - 1}<\omega_{j}\leq\omega, \\
			& \quad \text{ with }  k = \frac{\phi_{N+1} (\omega_k) + \omega_k}{\pi}.
		\end{split}
\end{equation}
Here $k\in[1,j]$ and it is an integer. Next, we derive the integrated DOS
from the total number of states.

\subsection{Integrated density of states}
The general form of the integrated DOS normalized over the number of sites is
\begin{equation}
      N_I(\omega) = \int^{\omega}_{-\infty} \rho(\omega') \rmd \omega'.
\end{equation}
Here $\rho(\omega)$ is the density of state (DOS).
In QW, particle-hole symmetry is present~\cite{Kitagawa2012},
so quasi-energy $\omega$ is symmetric with respect to 0. There are an
equal number of positive and negative quasi-energy states so that $N_I(0)=0.5$.

As shown in the previous section, the total number of states between
quasi-energies $0$ and $\omega$ is $j$,
and
\begin{equation}
   \label{IDOS_j}
   j=[(\phi_{N+1}(\omega)+\omega)/\pi],
\end{equation}
where $[x]$ denotes the largest integer
less or equal to $x$. So in our case,
\begin{equation}
   \label{IDOS_descrite}
   N_I(\omega)-N_I(0)=\frac{j}{N+1}.
\end{equation}
Now we need to evaluate $j$.

As shown in Eq.~\eqref{QW_newTMFII}, $\bv_{n+1}$ can be obtained from
$\bv_{n}$ after the operation $\SC_n\cdot \widetilde{R}$. The initial
vector $\bv_i$ will experience totally $N+1$ operations to reach the
final vector $\bv_f$. To see this, we add a matrix $\SC_{N+1}$ with $\vt_{N+1}=0$
to the right of Eq.~\eqref{RecurRelaBN3}. It is the identity matrix
so that Eq.~\eqref{RecurRelaBN3} holds. From $\bv_i$ to $\bv_f$,
the vector has passed many quadrants. We can define $N_q$ to
be the number of operations required
for the vector to leave the $q$-th quadrant since entering it.
Obviously, the summation of all the $N_q$ equals to $N+1$:
$\sum N_q = N+1$.

From $\bv_i$ to $\bv_f$, the vector rotates totally by an angle about $j\pi$ after $N+1$
operations (see Eq.~\eqref{IDOS_j}) 
so the number of quadrants passed is $2j$ and
\begin{equation}
      \sum^{2j}_{q=1} N_q = 2j\left(\frac{1}{2j} \sum^{2j}_{q=1} N_q \right)=2j\avg{N_q} = N+1.
\end{equation}
Hence, we have this formula~\cite{Eggarter1978},
\begin{equation}
       \label{DOS_I}
       N_I(\omega)-N_I(0)=j/(N+1)=\frac{1}{2\avg{N_q}},
\end{equation}
and $\avg{N_q}$ is the average number of operations required to pass
one quadrant since entering it. Equation~\eqref{DOS_I} resembles Eq.~$(21)$
in the paper by Eggarter and Riedinger~\cite{Eggarter1978}.  Though we approach
the DOS through counting the number of states like what was done in Ref.~\cite{Eggarter1978},
we are able to achieve this step by first introducing the transfer matrix approach when
analyzing the spinors in our QW model. More importantly, because the above expression for counting the number of states
is similar to that in Ref.~\cite{Eggarter1978}, we can now analogously
derive the DOS near $\omega=0$.


\subsection{Derivation of the DOS} 
\label{sec:DeriveDOS}
In the previous subsection, the integrated DOS is derived in Eq.~\eqref{DOS_I}, but
with one parameter $\avg{N_q}$ to be determined (which represents the average number of
operations required to pass one quadrant). Without loss of generality, we consider the first quadrant.

Let $z_n \equiv \cot{\phi_n}$. From Eq.~\eqref{RelaTan} we have
\begin{equation}
     \label{RelaZ}
     z_{n+1} = z_n \frac{1-(\tan\omega)/z_n}{1+z_n\tan\omega} \tan^2\vt_n.
\end{equation}
We define $u_n \equiv \ln z_n$ for $z_n\neq 0\text{ or }\infty$.
When
\begin{equation}
      \label{interval}
      \tan\omega \ll z_{n} \ll (\tan\omega)^{-1},
\end{equation}
one approximately has
\begin{equation}
     \label{randW}
     u_{n+1} \approx u_n + \ln\left(\tan^2\vt_n\right) .
\end{equation}
Since $\vt_n$ is taken randomly from this interval $[\pi/4-\tw,\pi/4+\tw]$,
we can conclude that $u_n$ executes a random walk~\cite{Eggarter1978}.
One may notice that the fraction factor in Eq.~\eqref{RelaZ}
is always smaller than 1 for positive $z_n$, so the random walk in Eq.~\eqref{randW}
is accompanied with a small negative drift. However, if the vector falls in the
second quadrant, the fraction factor will be always larger than 1, such that
the random walk has a small positive drift.
The two drifts cancel each other approximately.

When $u_n$ approaches the endpoints of the interval in~\eqref{interval},
the approximation in~\eqref{randW} no longer holds. Here we  analyze the situations
upon approaching the endpoints to show that they are similar to the situations analyzed in Ref.~\cite{Eggarter1978}.
If this is true, then the derivation there can be adopted here without much modification.

For $z_n \approx (\tan\omega)^{-1}$ (approaching the large $z_n$ limit), then $z_{n+1}\approx (1/2) z_n \tan^2\vt_n$
according to Eq.~\eqref{RelaZ}. The net shrinking factor (1/2) in this expression indicates that
$z_{n+1}$ will not keep growing. So $u_\text{max} = -\ln\tan\omega$ can be
considered as the reflection barrier as in Ref.~\cite{Eggarter1978}. We can also
view the reflection as the manifestation that the vector can never cross
$x$-axis clockwise (see Sec.~\ref{sec:DOS}.).

In the other extreme where $z_n \approx \tan\omega$ (approaching the small $z_n$ limit), the numerator
in Eq.~\eqref{RelaZ} will be much smaller than 1 so that $z_{n+1}<<z_n$,
indicating a sharp decrease in $z_n$. Once $z_n$ gets slightly below
$\tan\omega$, $z_{n+1}$ will be negative, indicating that the vector
moves into the second quadrant. So this boundary $u_\text{min} = \ln\tan\omega$
can be called an absorbing barrier~\cite{Eggarter1978}.
The vector passes positive $y$-axis counterclock-wise (see Sec.~\ref{sec:DOS}).

With all these, a mapping between our disordered QW model and the TBM with ODD
is established regarding all the system parameters. Specifically,
our Eqs.~\eqref{DOS_I}, \eqref{RelaZ} and \eqref{interval} resemble Eqs.~(21), (18) and (19)
in Ref.~\cite{Eggarter1978}, and the reflection and absorbing barriers
are similar, too. Further borrowing the method in Sec.~III of Ref.~\cite{Eggarter1978}, we directly
find $\avg{N_q}$
\begin{equation}
       \avg{N_q}  = \frac{4\ln^2\tan\omega}{\sigma^2}, \text{ with } \ \
			 \sigma^2  \equiv  2\expct{(\ln\tan^2\vt)^2}.
\end{equation}
Using Eq.~\eqref{DOS_I}, we obtain the integrated DOS,
\begin{equation}
      \label{DOS_I_f}
      N_I(\omega) = \frac{1}{2}\left(1+\frac{\sigma^2}{4\ln^2\tan\omega}\right),
\end{equation}
and then the DOS,
\begin{equation}
      \label{DOS_analytical}
      \rho(\omega) = \frac{\rmd N_I}{\rmd \omega} 
			             \approx -\frac{\sigma^2}{4} \frac{1}{\omega \ln^3 \omega }.
\end{equation}

To conclude, we have shown that our disordered QW model possesses the physics of
ODD.  It is for this reason that, quite remarkably, the derivation
of DOS for our QW model resembles to that in the original TBM with ODD~\cite{Theodorou1976,Eggarter1978}.
To make this connection between our QW model and the TBM with ODD clear is
the main contribution of this section.  We highlight the two crucial steps: (i)
linking the ``spinor'' components of the eigenstate through the transfer matrices,
and (ii) the interpretation of the eigenstate as a vector moving in the complex plane
when counting the number of states.


The localization length for quasi-energies around 0 can be derived in a similar
way~\cite{Eggarter1978} and the result is:
\begin{equation}
    \label{QW_LL}
        \ell^{-1}(\omega) \approx -\frac{\sigma^2\ln\omega}{4\ln^2\tan\omega} \approx -\frac{\sigma^2}{4\ln\omega}.
\end{equation}
Equation~\eqref{QW_LL} shows that
the localization length diverges as $\omega$ approaches 0, which is consistent
with the previously mentioned fact that the state with $\omega=0$ is delocalized.
\subsection{Numerical analysis of the DOS}
The derivation of DOS in Sec.~\ref{sec:DeriveDOS} involves some approximations, so we need numerical
simulations to check the analytical results. Specifically, we use
Eqs.~\eqref{RelaTan}, \eqref{IDOS_j} and \eqref{IDOS_descrite} to obtain
the integrated DOS numerically, and then compare our numerics with the analytical expression
given by Eq.~\eqref{DOS_I_f}. Given one disorder realization and one quasi-energy $\omega$,
we use the recursive relation in Eq.~\eqref{RelaTan} to obtain $\phi_{N+1}$,
and then it is substituted into Eq.~\eqref{IDOS_j} to obtain $j$, and
finally we get $N_I(\omega)$ through Eq.~\eqref{IDOS_descrite}.
Note that a randomly chosen $\omega$ may not be an actual quasi-energy value
associated with a particular disorder realization. However, if the system is sufficiently large, the quasi-energy values
will cover the vicinity of 0 quite densely. For this reason, a randomly chosen $\omega$ will not
cause noticeable error in terms of the counting of states.
\begin{figure}[!ht]
\centering
   \includegraphics[width=0.9\linewidth]{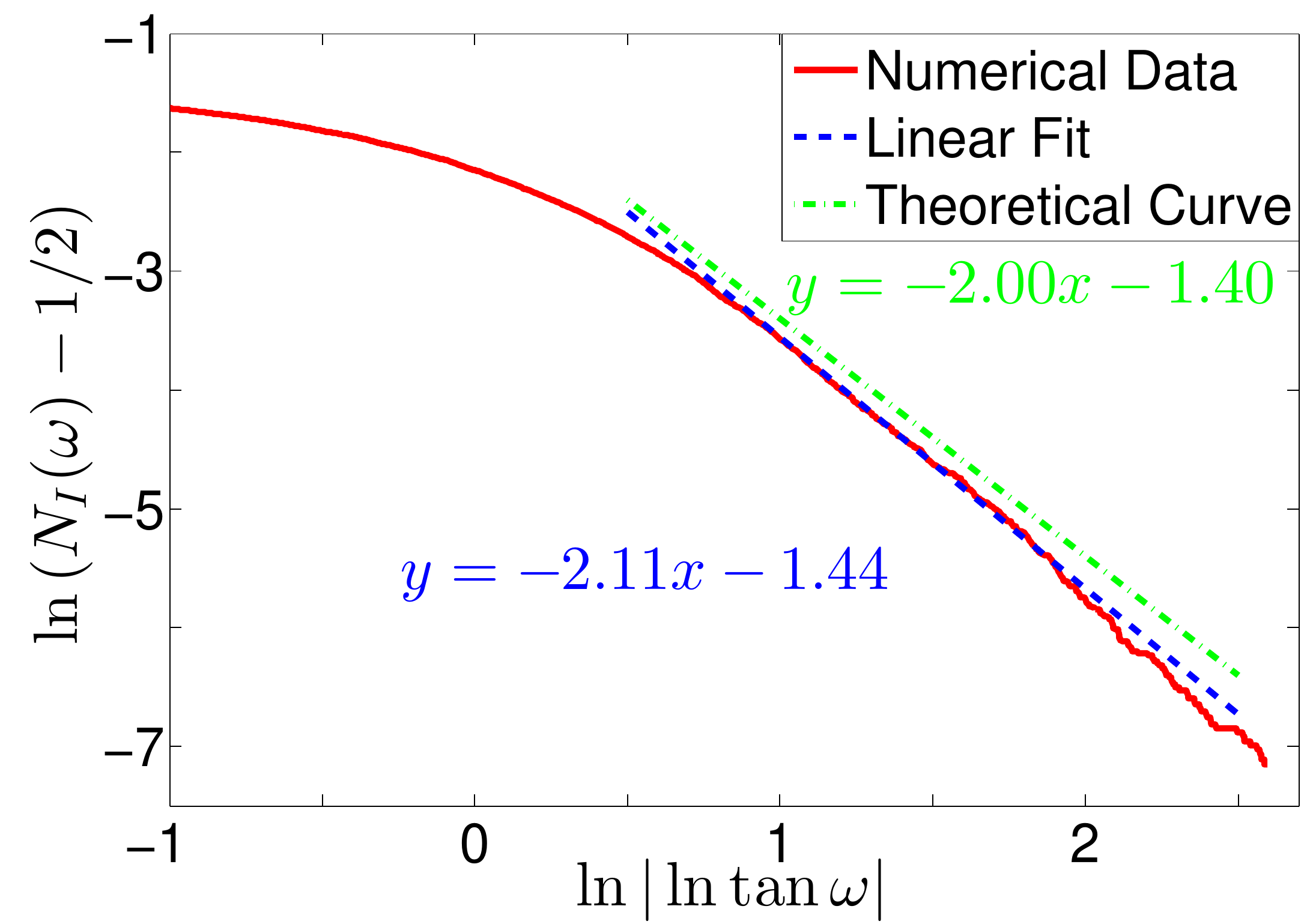}
   \caption[Relation between integrated DOS and quasi-energy $\omega$.]
	  {(Color online) Relation between integrated DOS and quasi-energy $\omega$, shown
via $\ln\left( N_I(\omega) - \frac{1}{2} \right)$ as a function of $\ln |\ln\tan\omega|$.
		The QW chain is of size $N=3\times 10^4$.
		The (red) solid line is from direct numerical calculations, the (blue) dashed line is a linear fit,
		and the (green) dash-doted line is our theoretical curve. The linear fit is applied to
		the domain $\ln |\ln\tan\omega|\in [1,2]$, corresponding to the quasi-energy domain
		$\omega\in[6.18\times 10^{-4},6.60\times 10^{-2}]$.
		}
   \label{fig:QW_IDOS}
\end{figure}

The analytical relation between $N_I(\omega)$ and $\omega$ is given by
Eq.~\eqref{DOS_I_f}. Alternatively,
\begin{equation}
      \label{DOS_I_fNum}
      \ln\left( N_I(\omega) - \frac{1}{2} \right) = \ln \frac{\sigma^2}{8} - 2\ln |\ln\tan\omega|.
\end{equation}
Figure~\ref{fig:QW_IDOS} depicts $\ln\left( N_I(\omega) - \frac{1}{2} \right)$ as a function
of $\ln |\ln\tan\omega|$ to check this theoretical prediction. The theoretical intersection on the $y$ axis
is $\ln \frac{\sigma^2}{8}\approx -1.40$ and the slope of the curve is $-2$. Our numerical results agree with
theory well in the main domain of our interest. However, for $\omega$ larger than $e^{-e}\approx 0.066$
(equivalently, $\ln |\ln\tan\omega|<1$),
theoretical results deviate from the numerical data, implying the failure of the analytical approximations
made in Sec.~\ref{sec:DeriveDOS}. This is expected as a too large $\omega$ leads to errors in
Eq.~\eqref{interval} and then in Eq.~\eqref{randW}.
In the case of $\omega<e^{-e^2}\approx 6.18\times 10^{-4}$
(equivalently, $\ln |\ln\tan\omega|>2$), the system size $N$ is no longer large enough for a reliable statistical analysis,
so the corresponding numerical results also start to deviate from our theoretical predictions.

\subsection{A numerical study of the self-correlation of delocalized states}

\label{sec:QW_analy_corr}
Here we numerically check whether the average two-point correlation of a
delocalized state with $\omega=0$ decays polynomially.
We use many realizations of disorder to obtain an average correlation function.
This is different from our previous calculations where only a single
realization of disorder is needed.
Analytically, assuming that a dimensionless product of disorder strength and two-point
separation is much larger than unity~\cite{Shelton1998},  the correlation exponent is shown to be $-3/2$.
This theoretical prediction is checked here by use of
Eqs.~\eqref{QW_TMM_diag} and \eqref{QW_TMM_diagII}, which depicts the eigenstate structure of our disordered QW model.

\begin{figure}[!ht]
\centering
   \includegraphics[width=0.9\linewidth]{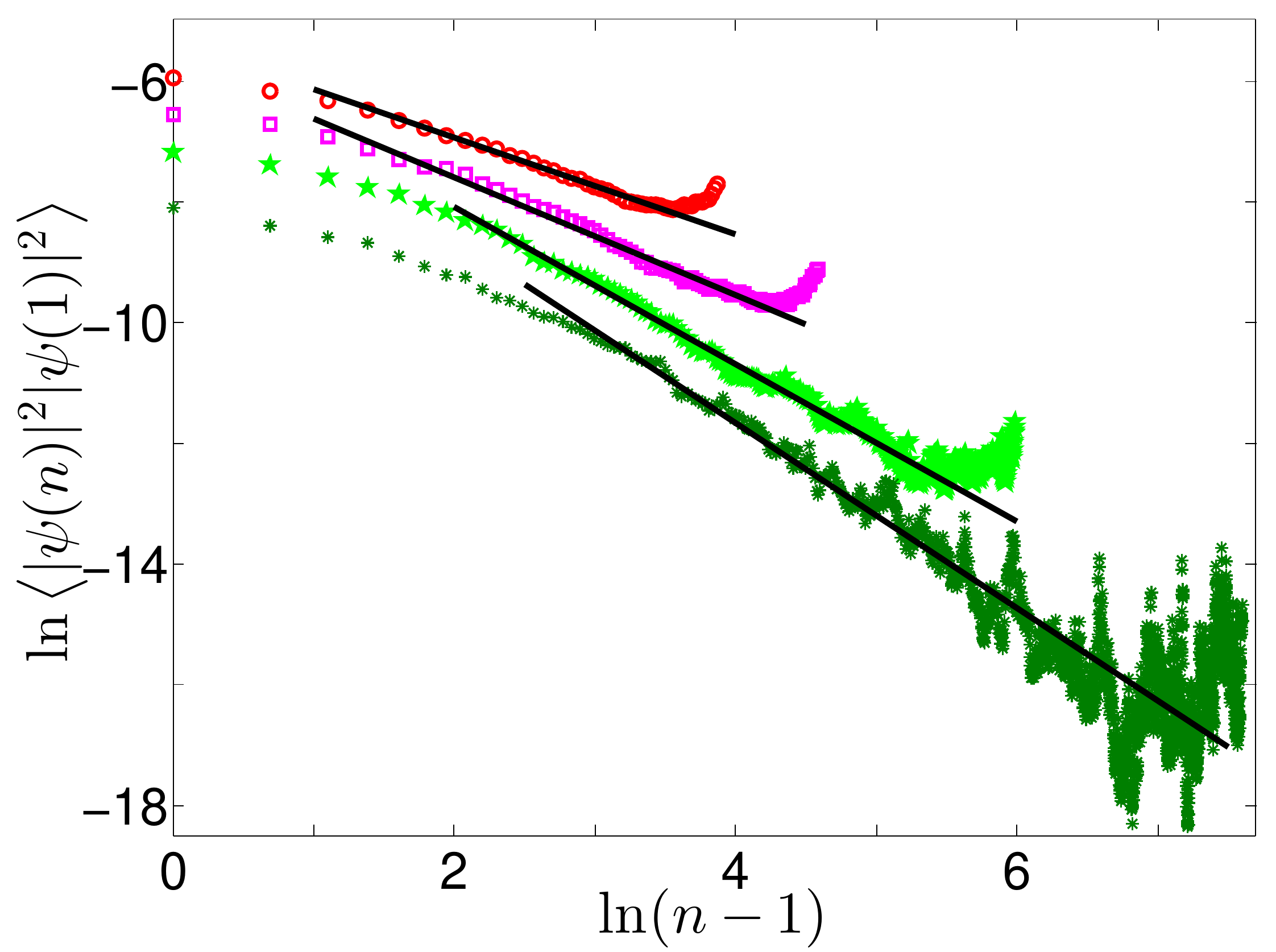}
   \caption[Dependence of correlation on different system size given fixed disorder strength.]
	  {(Color online) Dependence of correlation
   on the system size with the disorder strength fixed, as shown by
      $\ln\expct{|\psi(n)|^2|\psi(1)|^2}$ versus $\ln(n-1)$,
		averaging over $2000$ disorder realizations. Here
		$|\psi(n)|^2$ is the probability of the wave function at site $n$, and
		$\expct{|\psi(n)|^2|\psi(1)|^2}$ is the averaged two-point correlation, with one point
		fixed to be the site $1$. From top to bottom,
		the system size is set to be $N=50$, $100$, $400$ and $2000$
		respectively, and the linear fitting curves have slopes $-0.80$, $-0.98$,
		$-1.31$ and $-1.53$. The disorder strength is fixed to be $\tw=0.4$.		
		}
   \label{fig:QW_Exp_TM_Corr_N50_2000_w0_4}
\end{figure}
\begin{figure}[!ht]
\centering
   \includegraphics[width=0.9\linewidth]{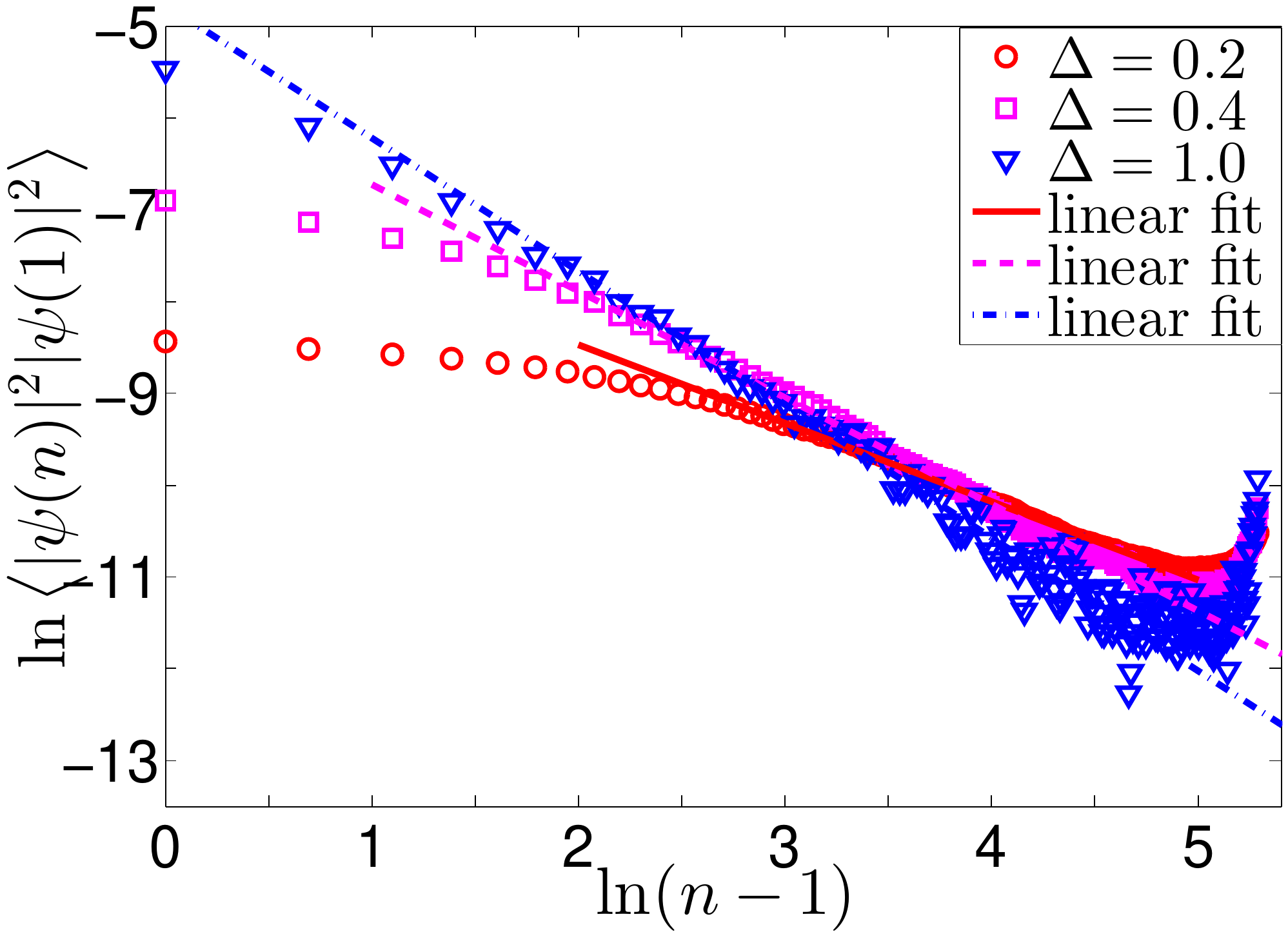}
   \caption[Dependence of the correlation on different disorder strength given fixed system size.]
	  {(Color online) Dependence of correlation on disorder strength with the system size fixed, as shown
 by  $\ln\expct{|\psi(n)|^2|\psi(1)|^2}$ versus $\ln(n-1)$,
		averaging over $10000$ disorder realizations.
		$|\psi(n)|^2$ is the probability of wave function at site $n$.
		System size $N+2=202$.
		Symbols circle, rectangle, and triangle represent
		$\tw=0.2$, $0.4$ and $1.0$
		respectively, and the linear fitting curves have slopes  $-0.86$, $-1.16$,
		and $-1.55$.
		}
   \label{fig:QW_Exp_TM_Corr_N200_w0_2_1_4}
\end{figure}

In Fig.~\ref{fig:QW_Exp_TM_Corr_N50_2000_w0_4}, the disorder
strength is set to be $\tw=0.4$, and the system size varies from $N+2=52$ to
$2002$. When the two-point separation increases,
the correlation exponent increases from $0.8$ to $1.5$ and stays almost stable at $1.5$.
Figure \ref{fig:QW_Exp_TM_Corr_N200_w0_2_1_4} shows how the
correlation varies with the disorder strength. The general observation
is that increasing the
disorder strength will increase the correlation exponent
but the exponent again tends to saturate around $-3/2$. These numerical results
are consistent with the early theoretical prediction of ODD~\cite{Balents1997,Shelton1998}.
However, we point out that if $N$ and $\tw$ are too large, the statistical fluctuations become more pronounced
due to our limited number of realizations of disorder.

\section{Experimental Preparation of the 0-mode in disordered QW}
\label{sec:QW_expt}
It is now clear that when the disordered local rotation angle variables $\theta_n$
fluctuate around zero (i.e., $\tilde{\theta}=0$ in Eq.~\eqref{QW_BulkAngle}), then
the 0-mode (eigenstate with $\omega=0$) in our disordered QW model
reflects the physics of ODD.  However,  if $\tilde{\theta}\ne 0$, then the corresponding
0-mode becomes unrelated to ODD physics.
For example, if $\theta_n$ slightly fluctuates around $\pi/2$, then the 0-mode will still be
highly localized around the sites $0$ and $1$, with negligible proportion in all other
sites. 

The 0-mode with $\tilde{\theta}=0$ is in general delocalized and hence it is
hard to prepare in experiments.  To address this issue, we note that the highly localized 0-mode associated with
$\tilde{\theta}=\pi/2$ is a good starting point.  We propose to connect this localized 0-mode
with our target 0-mode possessing ODD physics by an adiabatic protocol~\cite{Dranov1998,Tanaka2011,hailongpaper}. That is,
by slowly tuning the value of $\tilde{\theta}$ from $\pi/2$ to 0, we may reach our target $0$-mode from
the localized 0-mode.

Consider then a conventional adiabatic evolution protocol, through which
the parameters $\theta_n$ in the QW operator $U$ are tuned slowly.
Note, however, that the boundary rotation angles $\theta_0$ and $\theta_{N+1}$
must be fixed to ensure the conservation of probability inside the QW chain.
An adiabatic process reflecting this constraint is as follows.
At first, the system is set as $\theta_0=-\pi/2$, $\theta_1=\theta_{N+1}=\pi/2$
and $\theta_n=\pi/2+\delta_n$ with $n\in [2,N]$ and $\delta_n$ being random angle fluctuations.  The mean value of $\delta_n$ over $N$ sites is denoted $\bar{\delta}$.
The initial state of the QW model is prepared with entries $\beta_0=\alpha_1=1/\sqrt{2}$ and all other entries 0.
It can be easily checked that this initial state is precisely the 0-mode of the system (note that $\theta_1$ is chosen to be $\pi/2$).
Then, we slowly reduce $\theta_n$ during the QW process, until $\theta_n = \delta_n$.
To be more specific, the proposed adiabatic protocol can be achieved by introducing a slow time dependence to $\tilde{\theta}$
in Eq.~\eqref{QW_BulkAngle}, i.e.,
\begin{equation}
    \label{QW_adia_process}
     \theta_n (t) =\tilde{\theta}(t) + \delta_n,
\end{equation}
with $n\in[1,N]$ denoting the bulk-site index, $\delta_1=0$, and $\tilde{\theta}(t)$ to be further specified below.

The QW mapping operator $U$ associated with $\theta_n(t)$ is denoted as $U(t)$.
The initial state $\ket{\psi(0)}$ is localized at the first two sites, with $U(0)\ket{\psi(0)} = \ket{\psi(0)}$.
The time-evolving state at time $t$ is denoted $\ket{\psi(t)}$, obtained by
\begin{equation}
     \label{AdiabaticEvolv}
     \ket{\psi(t)} = U(t)\cdot U(t-1) \cdots U(1)\cdot U(0) \ket{\psi(0)}.
\end{equation}
For the sake of comparison between the time evolving state $\ket{\psi(t)}$ and our target 0-mode state,
we define the exact zero-quasienergy eigenstate of $U(t)$ as $\ket{\psi^0(t)}$ (with $U(t)\ket{\psi^0(t)} = e^{i\cdot 0} \ket{\psi^0(t)}$).  Numerically we can directly diagonalize $U(t)$ to get $\ket{\psi^0(t)}$.   Our hope is to
reach $\ket{\psi^0(t)}$ through the time evolving state $\ket{\psi(t)}$ emerging from our adiabatic protocol.
Indeed, the adiabatic theorem~\cite{Dranov1998,Tanaka2011,hailongpaper} states that
$\ket{\psi(t)} \approx \ket{\psi^0(t)}$ if the adiabatic conditions are fulfilled.

We have numerically simulated the process depicted in
Eq.~\eqref{AdiabaticEvolv}, and then compare $\ket{\psi(t)}$ with $\ket{\psi^0(t)}$.
Their overlap probabilities $|\ip{\psi(t)}{\psi^0(t)}|^2$ versus $t$ is plotted to check the
performance of a certain specific protocol. In the following,
by specifying $\tilde{\theta}(t)$ differently, we examine
two protocols to realize the adiabatic process and hence the preparation of the target 0-mode state that reflects the physics of
ODD.

\subsection{Tuning $\tilde{\theta}$ at a constant rate}
\label{sec:QW_uniform}
\begin{figure}[!ht]
\centering
   \includegraphics[width=\linewidth]{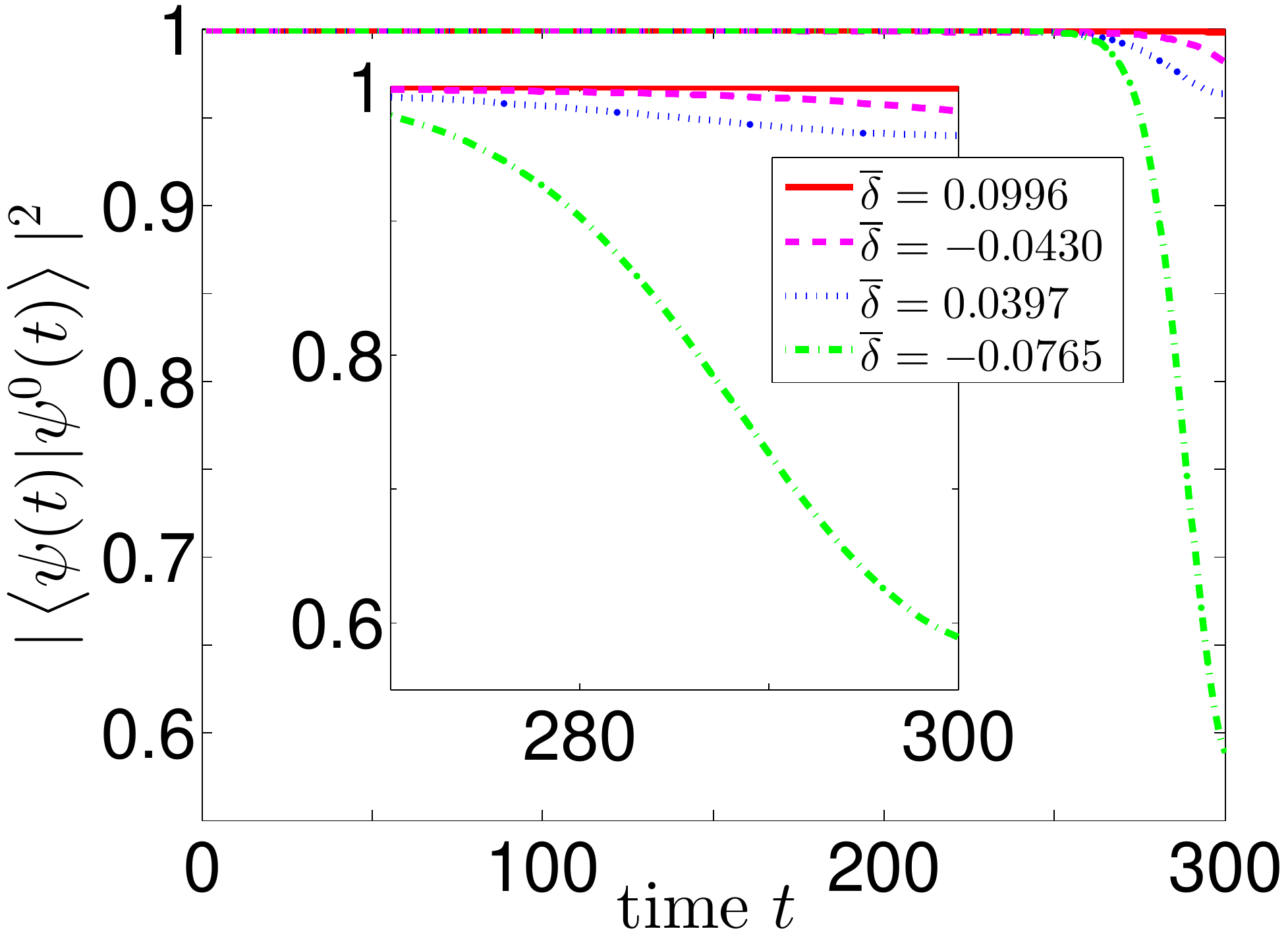}
\caption[Relations among gap size, averaged bulk $\theta_n$ and evolving time $t$.]{
    (Color online) Overlap probability between the actual time evolving state $\ket{\psi(t)}$ and instantaneous 0-modes
			$\ket{\psi^0(t)}$ for 4 realizations of disorder in numerical experiments.
		The inset is a magnified
		view of the tail part.
		The (red) solid, (pink) dashed, (blue) dotted and (green) dash-dotted lines represent 4 different
		realizations of disorder with different $\avg{\delta}$ (shown on the figure panel).
		The disordered chain has totally $N+2=20$ sites, with the disorder strength given by $\tw=0.7$.
		}
\label{fig:QW_1stProtocol}
\end{figure}

In this case we decrease the bulk $\theta_n$ at a
constant rate with respect to the evolution time.
Specifically, $\tilde{\theta}(t)$ in Eq.~\eqref{QW_adia_process} is given by
\begin{equation}
     \tilde{\theta} (t) = \tilde{\theta} (0) - r t,
\end{equation}
where $t=0,1,2,\ldots,T$ is the evolution time, $r=\tilde{\theta} (0)/T$ is the constant
decreasing rate, and $\tilde{\theta} (0)=\pi/2$.  The obtained state fidelity
$|\ip{\psi(t)}{\psi^0(t)}|^2$ versus $t$ is plotted in Fig.~\ref{fig:QW_1stProtocol}.

Figure~\ref{fig:QW_1stProtocol} shows that for some realizations of disorder, the fidelity
near the final stage of the evolution decreases significantly. The difference seems to be
related to $\bar{\delta}$, the actual mean value of the random fluctuations $\delta_n$ in a particular realization
of disorder.  In particular, the realization with $\avg{\delta}\approx -0.076$
(green dash-dotted line) has a final fidelity below $0.6$.
To understand this, we investigate the gap between the 0-mode and its neighboring mode,
which is found to decrease with $t$.
When $\tilde{\theta}(t)$ gets close to 0, the 0-mode is not well separated from the bulk modes, and
the gap becomes quite small. Compared with other three realizations, the
realization with $\avg{\delta}\approx -0.076$ has a gap size of approximately half
of others from $t\approx 250$ to $300$, so this small gap has caused the most pronounced nonadiabatic transitions.
To confirm this, we increase the
total evolution time and indeed a better performance can be obtained (see
Fig.~\ref{fig:QW_adia_expon_probVSt_allTs} presented later).
By contrast, for other realizations in Fig.~\ref{fig:QW_1stProtocol}, the final fidelity is
high (above $0.95$), an indication of good performance due to the associated relatively large gaps.
To summarize, the performance of this adiabatic protocol is determined by the
total evolution time $T$ and the gap size in the final evolution stage. One can always
improve the performance by increasing $T$. In contrast, the gap size is sensitive to
the details of an actual realization of disorder.  As an observation from our numerical results,
cases with a negative $\avg{\delta}$  tend to
have a smaller gap size around the final evolution stage than cases with a positive $\avg{\delta}$.

\subsection{Tuning $\tilde{\theta}$ exponentially}
\label{sec:QW_adia_expon}


To understand our motivation of this alternative protocol, we first discuss the gap size of the
clean system,  where the bulk $\theta_n$ is uniform (i.e., $\theta_n=\tilde{\theta}$). In this case,
two quasi-energy bands emerge and the dispersion relation
is given by $\cos\omega = \cos\tilde{\theta} \cos k$~\cite{Kitagawa2012},
where $k$ is the quasi-momentum. The gap between the bands is $2\tilde{\theta}$ at $k=0$.
The 0-mode sits in the center of the band gap.
We are thus motivated
to design the following protocol by roughly assuming that the gap between the 0-mode and the bulk spectrum
is proportional to $\tilde{\theta}$:
\begin{equation}
     \label{QW_expon_adia}
      \frac{\rmd}{\rmd t} \tilde{\theta}(t) = - \lambda \tilde{\theta}(t).
\end{equation}
In this new protocol, the rate of change $\frac{\rmd}{\rmd t} \tilde{\theta} (t) \propto$
instantaneous gap $\propto$ instantaneous $\tilde{\theta}(t)$ .
As the gap decreases, the rate of change also decreases to keep the
process being sufficiently adiabatic.
Therefore $\tilde{\theta}$ is an exponential function of $t$,
\begin{equation}
    \label{QW_expon_adiaII}
     \tilde{\theta} (t) = \tilde{\theta}(0)e^{-\lambda t},  
\end{equation}
where $\lambda$ is the exponential decay rate of $\tilde{\theta}$.
Using this protocol, $\theta_n$ can be explicitly expressed as a function of $t$:
\begin{equation}
    \label{QW_expon_protocol}
     \theta_n (t)=\begin{cases}
		                  -\frac{\pi}{2}  &  n=0, \\
											\tilde{\theta} (t)  &  n=1, \\
											\tilde{\theta} (t) - \frac{N}{N-1}\tilde{\theta} (T)+ \delta_n & n\in [2,N], \\
											\frac{\pi}{2}   &  n=N+1.
		              \end{cases}
\end{equation}
Here $\frac{N}{N-1}\tilde{\theta} (T)$ is to make sure that
$\sum^{N}_{n=1} \theta_n(t) = \sum^N_{n=1} \delta_n$ at the final time $t=T$.
Note also that at site $n=1$, $\theta_1(0)=\tilde{\theta}(0)=\pi/2$, which
ensures that the initial 0-mode is the exact eigenstate of the QW propagator at time zero.
\begin{figure}[!ht]
\centering
   \includegraphics[width=\linewidth]{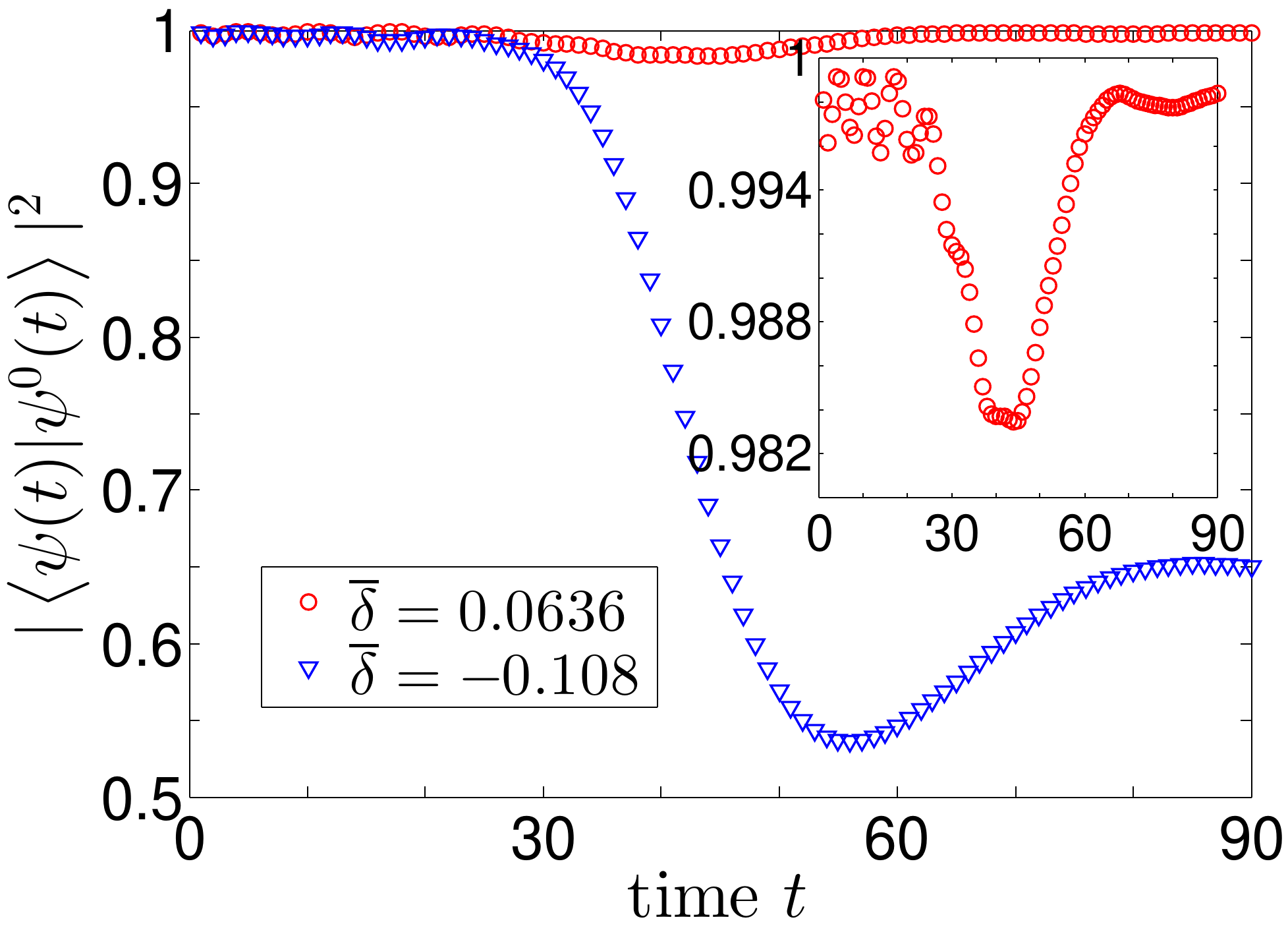}
\caption[Overlap probability between the actual time evolving state and and the instantaneous 0-modes.]{
      (Color online) Overlap probability between the actual time evolving state $\ket{\psi(t)}$ and instantaneous 0-modes
			$\ket{\psi^0(t)}$ for 2 different types of disorder realization. The chain has $N+2=20$
			sites, total evolution time $T=90$, disorder strength $\tw=0.7$, and the parameter in the exponential protocol
is characterized by $\lambda=0.0562$.
			(red) Circles are for a case with the averaged angular disorder $\avg{\delta}=0.064$ being positive,			
			with the overlap probability above 0.998 at the final time. The inset shows
			more details.
			(blue) Triangles for a case with the averaged angle disorder $\avg{\delta}=-0.108$ being negative.
			In this case, the final overlap probability is only around 0.65, which means
			that this protocol is still not working well with $T=90$.
			}
\label{fig:QW_adia_expon_probVSt}
\end{figure}

Figure~\ref{fig:QW_adia_expon_probVSt} shows the performance of this protocol.
For positive $\avg{\delta}$, the overlap probability at final time is quite high (above 0.998).
Interestingly, similar to the previous protocol in which we sweep $\tilde{\theta}$ at a constant rate,
the fidelity degrades
in cases of $\avg{\delta}<0$.  In addition, in some realizations of disorder, the gap size may be erratic
during the last stage of the adiabatic protocol, especially when $\avg{\delta}$ turns from positive to negative.
This explains the relatively poor performance
for the case with $\avg{\delta}=-0.108$ in Fig.~\ref{fig:QW_adia_expon_probVSt}.

\begin{figure}[!ht]
\begin{center}$
\begin{array}{c}
\begin{overpic}[width=65mm, height=!]{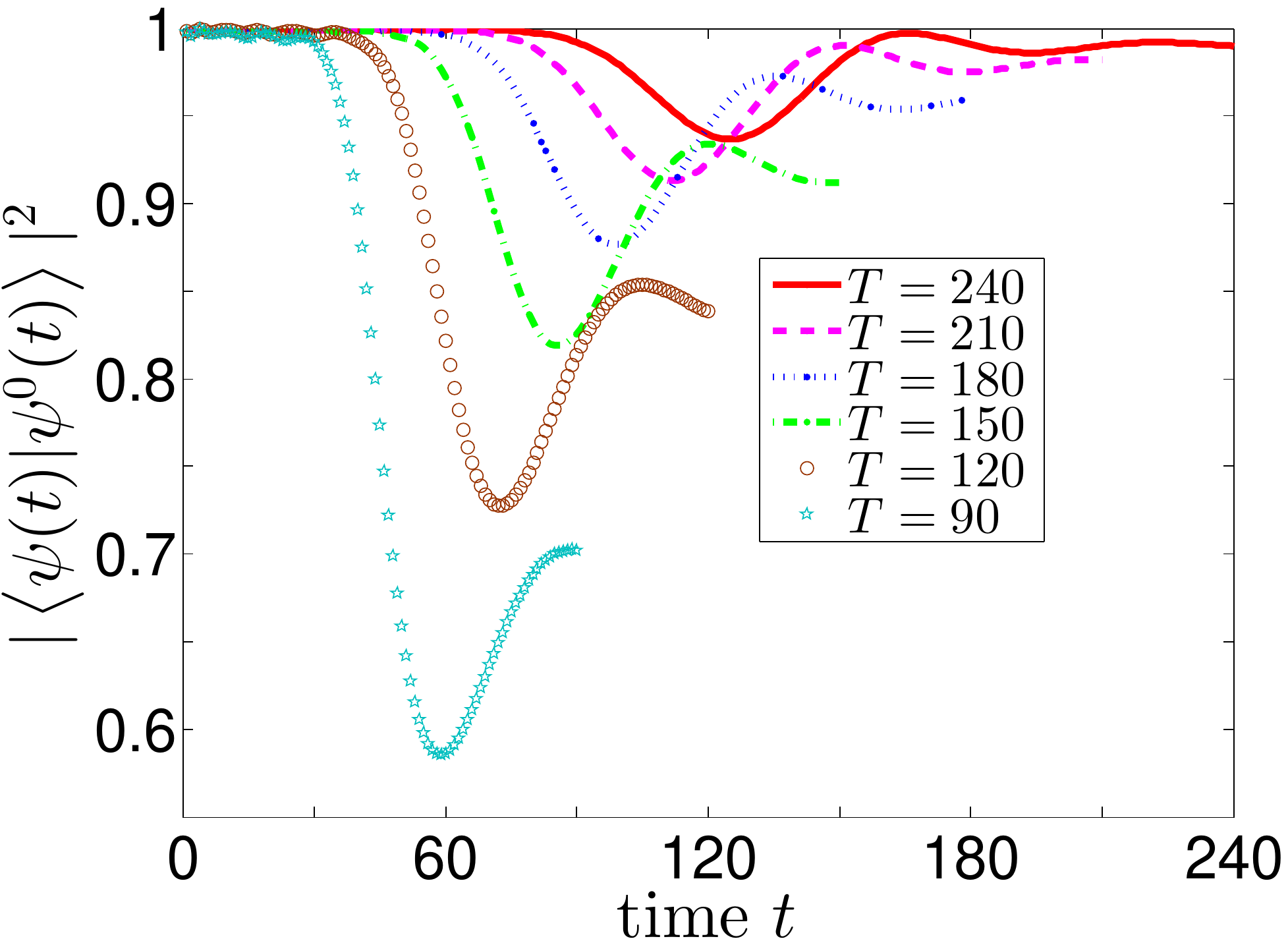}
\put (17,60) {\large$(a)$}
\end{overpic}\\
\begin{overpic}[width=65mm, height=!]{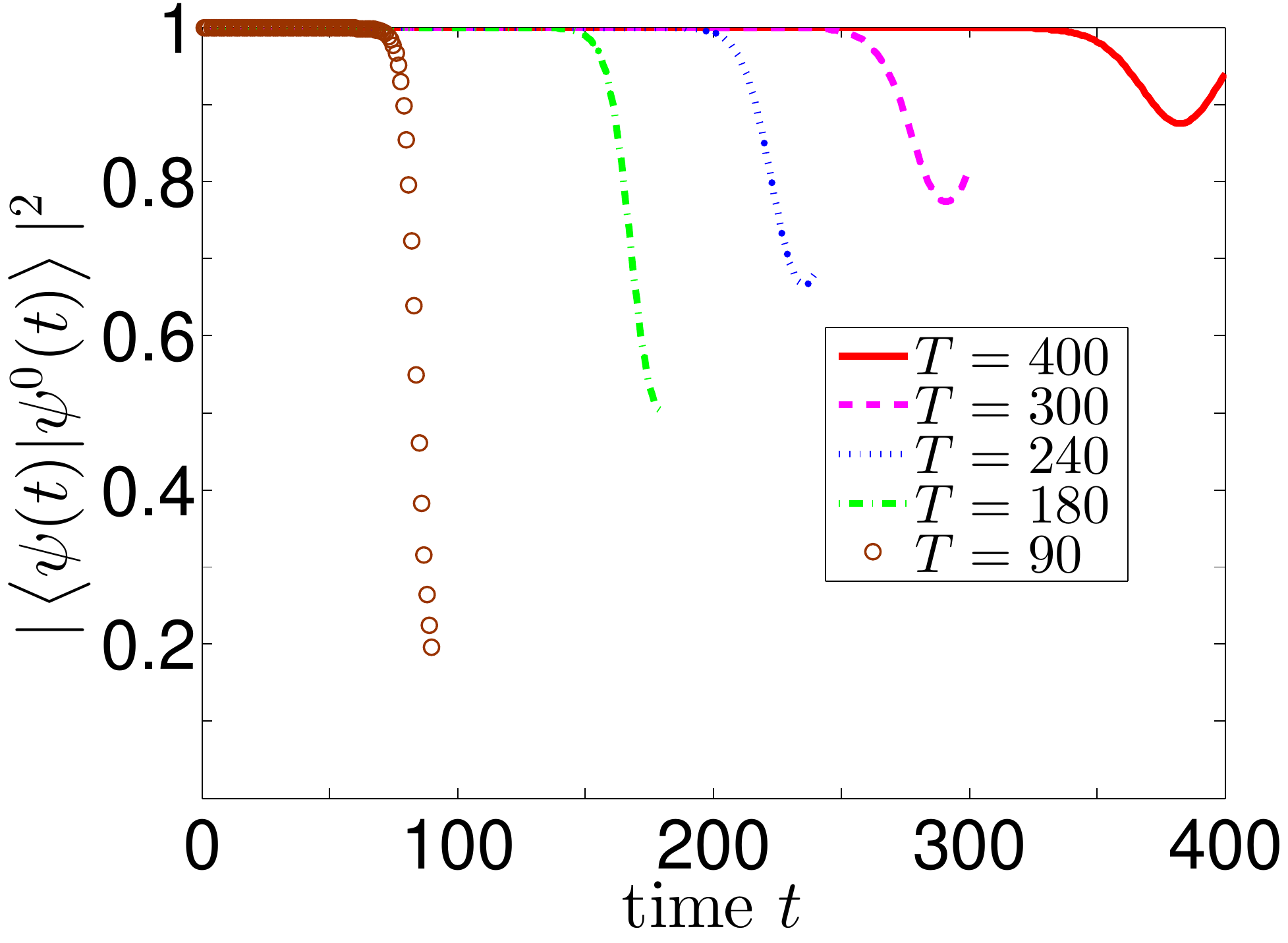}
\put (19,60) {\large$(b)$}
\end{overpic} \\
\end{array}$
\end{center}
   \caption[Performance of adiabatic processes with different $T$s.]
	  {(Color online) Overlap probability versus $t$ for different protocol duration $T$, for
		an exponential protocol $(a)$  (Eq.~\ref{QW_expon_adiaII}))
        and the previous constant-rate protocol $(b)$.
		In both protocols, the disorder realization is the same as the one with
		$\avg{\delta}=-0.1083$	in Fig.~\ref{fig:QW_adia_expon_probVSt}, and $N+2=20$, $\tw=0.7$.
		$(a)$ From top to bottom, $T$ equals
		$240$, $210$, $180$, $150$, $120$ and $90$.  The corresponding values of $\lambda$ is chosen to be
		$\lambda = -\ln(0.01/(\pi/2))/T$. $(b)$
		From top to bottom, $T$ equals $400$, $300$, $240$, $180$ and $90$.
        In both panels,
		a larger $T$ results in a better fidelity of the final state.
		However, the exponential protocol in general requires less time to achieve the same fidelity.
		}
   \label{fig:QW_adia_expon_probVSt_allTs}
\end{figure}

Nevertheless, we can further improve the fidelity by increasing the total
evolution time $T$ or decreasing $\lambda$ in our exponential protocol.
Panel $(a)$ of Fig.~\ref{fig:QW_adia_expon_probVSt_allTs} how fidelity changes with $T$. As a comparison,
in panel $(b)$ of Fig.~\ref{fig:QW_adia_expon_probVSt_allTs} we show the parallel fidelity vs $T$ if $\tilde{\theta}$ is swept
at a constant rate.  It is seen that overall, tuning $\tilde{\theta}$ exponentially as is done here is much better than
tuning $\tilde{\theta}$ at a constant rate.

\subsection{Correlation exponents in numerical experiments}
We have shown in the previous subsection how to prepare the 0-mode state possessing the physics
of ODD. Here we aim to show that states prepared in this manner can indeed manifest the
correlation exponent characteristic of ODD physics.  In doing so we need to
perform averaging over many realizations of disorder.  We use the exponential adiabatic protocol in our numerical experiment.
To benchmark our numerical experiments, we also analyze the correlation exponent using the exact
delocalized 0-mode state obtained from Eqs.~\eqref{QW_TMM_diag} and \eqref{QW_TMM_diagII}.

Before presenting our results,  we first discuss two minor issues.
The first is related to the fact that
the spinors represented in Fig.~\ref{fig:QW_setup} involve two different sites. That is,
In a real experiment, what is measured is likely the probability  at each site, whereas
in our analytical study, we treat $(\beta_{n-1}\ \alpha_{n})^\text{T}$ as one ``spinor''.  However, we find
that this difference has little effect on the correlation exponent. The other issue
is that we have fixed $\theta_1$ to be $\pi/2$ (hence not random) (see Sec.~\ref{sec:QW_expt} for details).
Again, it is checked that this does not affect our analysis.
\begin{figure}[!ht]
\begin{center}$
\begin{array}{c}
\begin{overpic}[width=65mm, height=!]{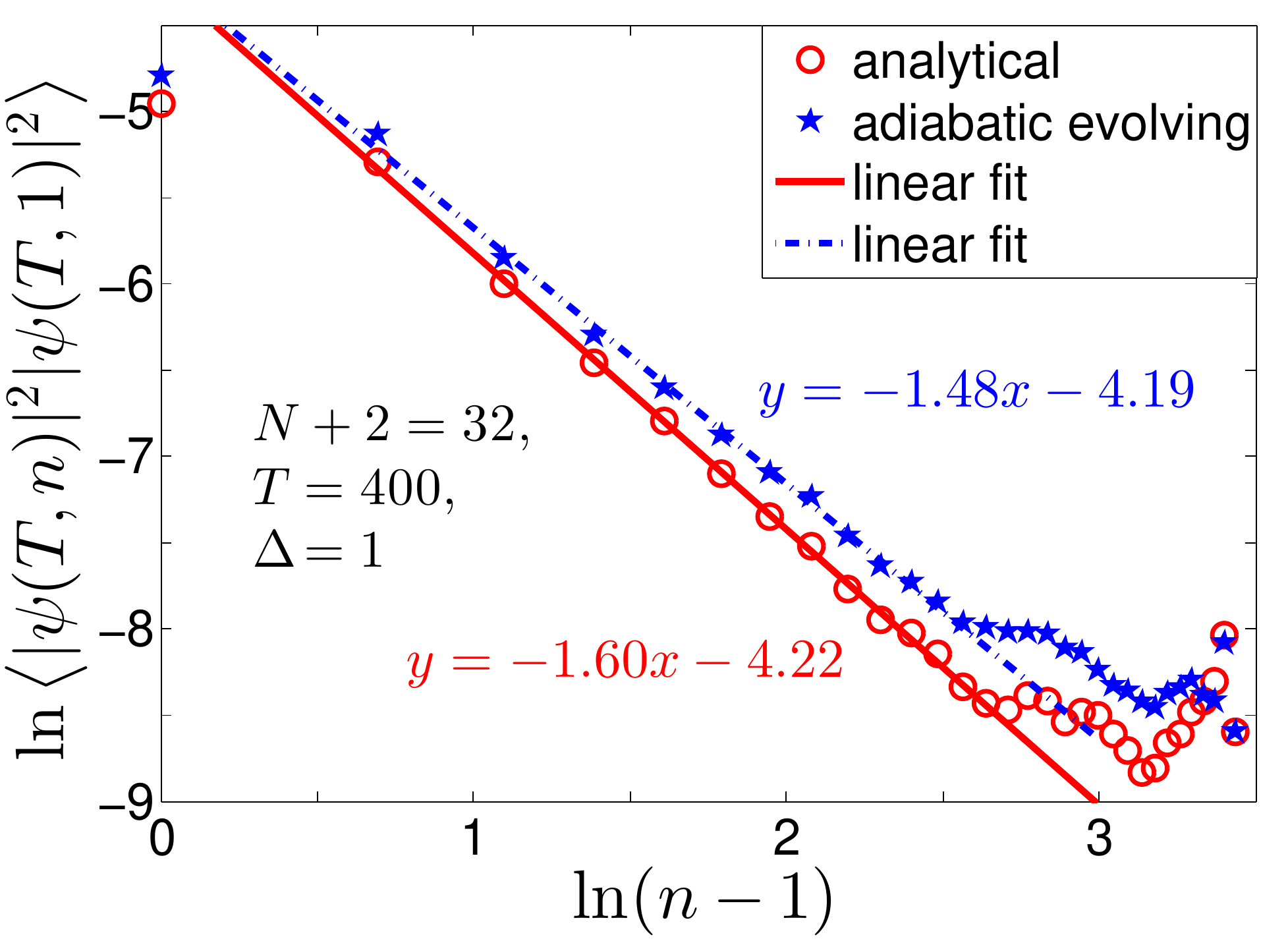}
\put (20,50) {\large$(a)$}
\end{overpic}\\
\begin{overpic}[width=65mm, height=!]{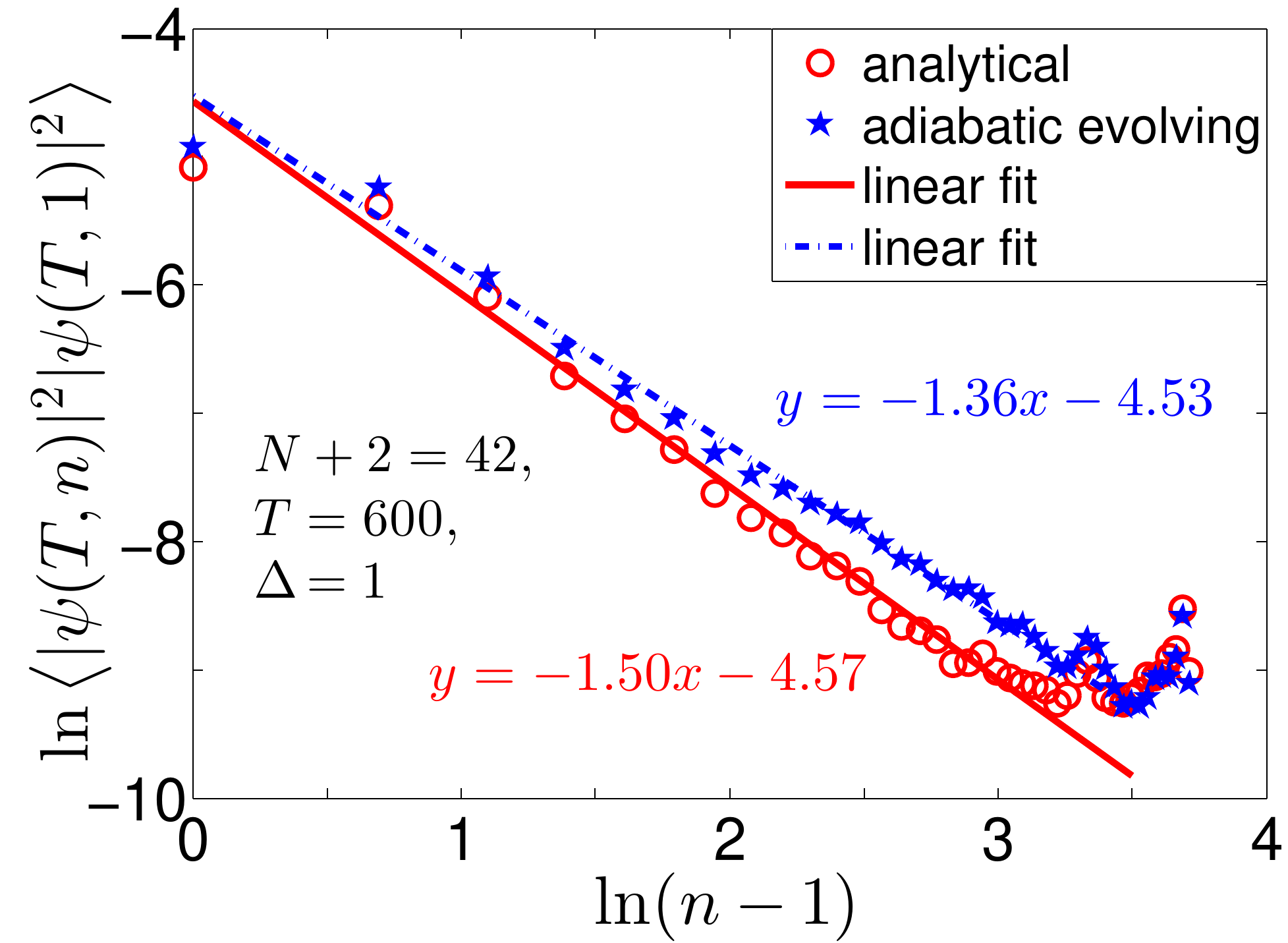}
\put (20,47) {\large$(b)$}
\end{overpic} \\
\end{array}$
\end{center}
\caption[$-3/2$ correlation exponents in numerical experiments.]{
      (Color online) Correlation function $\ln\expct{|\psi(T,n)|^2|\psi(T,1)|^2}$ versus $\ln(n-1)$,
			averaged over 1000 disorder realizations.
			$(a)$ disorder strength $\tw=1$, system size $N+2=32$;
			$(b)$ $\tw=1$, $N+2=42$. The total evolution time $T$
			is chosen to assure satisfactory fidelity in the adiabatic preparation
            of the 0-mode, with $T=400$ in panel $(a)$ and $T=600$ in panel $(b)$.
			In both panels (red) circles denote results from
			solving the 0-mode analytically; whereas (blue) stars denote
			results obtained from our adiabatic preparation of the 0-mode with the exponential protocol.
            Solid line and dash-dotted line are the associated linear fitting curves over a regime without much fluctuation.
            The slopes of the fitting curves reflect the correlation exponents.
			}
\label{fig:QW_Exp_Corr_Diff_Ns_Essen}
\end{figure}

We also note that the $-3/2$ correlation exponent was derived
under the assumption that the product of the dimensionless disorder strength and two-point
separation is much larger than unity~\cite{Shelton1998}. In real experiments,
the QW chain might not be long, so we are limited to relatively small two-point separation.
That means we should choose strong disorder strength to fulfill this
assumption. Figure \ref{fig:QW_Exp_Corr_Diff_Ns_Essen} presents our results from numerical experiments based
on an exponential adiabatic protocol starting from a highly localized state, as compared with a direct investigation
using the exact delocalized 0-mode states. For two different chain length, the two-point correlation exponents in our numerical experiments are found to be
$-1.48$ and $-1.36$, as compared with $-1.6$ and $-1.5$ obtained from pure theory.
Certainly, the agreement between these two sets of data can be further improved
if we further increase $T$.  The conclusion is that our adiabatic protocol applied to our disordered QW model
is also useful in the actual demonstration of
the two-point correlation characteristic of ODD physics.

For small systems with weak disorder, the analytical correlation
exponents are not available~\cite{Shelton1998}. To motivate experimental studies on this
matter, below we further exploit
our setup to investigate how the two-point correlation changes
with weak disorder strength $\tw$ and system size ($N+2$).

\begin{figure}[!ht]
\centering
   \includegraphics[width=\linewidth]{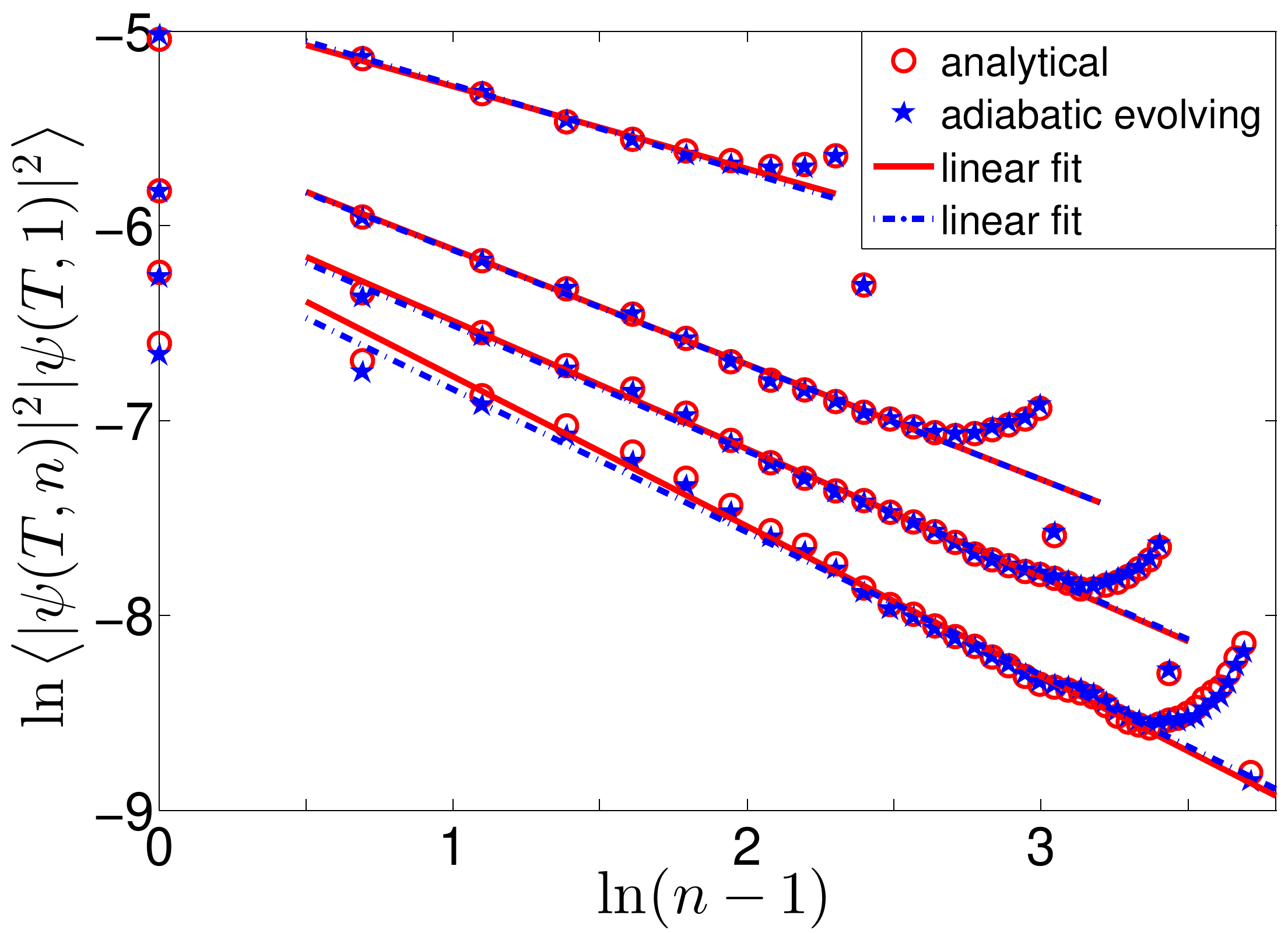}
\caption[Correlation exponents of systems with different size in numerical experiments.]{
      (Color online) Correlation functions with weak disorder in a QW model, shown via
          $\ln\expct{|\psi(T,n)|^2|\psi(T,1)|^2}$ versus $\ln(n-1)$,
			averaged over 1000 disorder realizations.
			The total evolution time $T$
			is chosen to make sure the adiabatic protocol can yield a satisfactory fidelity
            of the 0-mode state. For example, if $\tw$ or $N$ is
			increased, $T$ is increased also (See Sec.~\ref{sec:QW_adia_expon}).
			Here $\tw=0.4$, and from top to bottom, the system size is $N+2=12$,
			$22$, $32$ and $42$ respectively. The slopes of the curves fitting the results using the exact 0-mode
			(red solid line) are $-0.45$, $-0.59$, $-0.65$ and $-0.77$, whereas the
			slopes of the curves fitting the results arising from our adiabatic protocol
            (blue dash-dotted line) are $-0.42$, $-0.59$, $-0.66$ and $-0.73$ respectively.
			The symbols and the lines share the same meaning with those in
			Fig.~\ref{fig:QW_Exp_Corr_Diff_Ns_Essen}.
			}
\label{fig:QW_Exp_Corr_Diff_Ns}
\end{figure}

We choose 4 different system sizes with a fixed and weak disorder strength $\tw=0.4$.
In particular, we let $N+2=12$, $22$, $32$ and $42$. The results are shown
in Fig.~\ref{fig:QW_Exp_Corr_Diff_Ns}. For each case, we show statistical results obtained from
analytical treatment of the $0$-mode with disorder and from our exponential adiabatic protocol that starts from
an initial localized state.
The results obtained from such two totally different methods agree very well because
they yield almost the same slopes from the fitting straight lines, for all the four cases shown.
The good fitting by the straight lines indicates a polynomial behavior of the two-point correlation function, but
now with correlation exponents given by $-0.447$, $-0.588$, $-0.645$ and $-0.769$, for
$N=10$, $20$, $30$ and $40$, respectively.   These exponents are far from -3/2, but shows a tendency to approach -3/2
as the system size increases. Further increasing the value of $\tw$ also increases the magnitude of the correlation exponent. These results should be of experimental interest as well and invite further theoretical developments in studies of the physics of ODD.

\section{Summary}
\label{sec:QW_con}
To summarize, we have shown that the physics of
ODD can be investigated by a disordered QW model. The associated
exotic features in the delocalization and in the wavefunction correlation are derived and
numerically verified. Because the physics of ODD
is rarely cleanly observed in actual experiments, our results will
possibly motivate ongoing QW experiments as a new platform to study the physics of ODD.
To facilitate such efforts, we proposed and analyzed adiabatic protocols to prepare the
exotic delocalized $0$-mode state with good fidelity. Our numerical experiments show that the delocalized $0$-mode states
thus obtained can directly show the correlation exponent -3/2 in the regime predicted by existing theory. Our numerical experiments
also show that
much different correlation exponents emerge if the product of the system size and the disorder strength is relatively
small.

\appendix

\section{From Eq.~\texorpdfstring{\eqref{RecurRela}}{TEXT} to Eq.~\texorpdfstring{\eqref{RecurRelaBN3}}{TEXT}}
\label{sec:EqTrans}
Here we show how to derive Eq.~\eqref{RecurRelaBN3} from
Eq.~\eqref{RecurRela}.
Multiply both sides of Eq.~\eqref{RecurRela} with $e^{-i\frac{\omega}{2}}$, and decompose
$T_n$ using the following identity
\begin{equation}
   \begin{split}
     T_n & = \begin{pmatrix}
		               e^{i\omega} \sec\theta_n   &  -\tan\theta_n \\
							     -\tan\theta_n &  e^{-i\omega} \sec\theta_n
		        \end{pmatrix} \\
		 & \equiv  \begin{pmatrix}
		               e^{i\frac{\omega}{2}}   &  0 \\
							     0                       &  e^{-i\frac{\omega}{2}}
		        \end{pmatrix}
		        \begin{pmatrix}
		               \sec\theta_n   &  -\tan\theta_n \\
							     -\tan\theta_n  &  \sec\theta_n
		        \end{pmatrix}
						\begin{pmatrix}
		               e^{i\frac{\omega}{2}}   &  0 \\
							     0                       &  e^{-i\frac{\omega}{2}}
		        \end{pmatrix},
	 \end{split}
\end{equation}
then Eq.~\eqref{RecurRela} becomes
\begin{equation}
     \label{RecurRelaBN}
		\begin{split}
		 \tmatrix{e^{-i\frac{\omega}{2}}}{e^{i\frac{\omega}{2}}}
      & = c
		        \begin{pmatrix}
		               e^{i\frac{\omega}{2}}   &  0 \\
							     0                       &  e^{-i\frac{\omega}{2}}
		        \end{pmatrix}
						\begin{pmatrix}
		               \sec\theta_{N}   &  -\tan\theta_{N} \\
							     -\tan\theta_{N}  &  \sec\theta_{N}
		        \end{pmatrix} \cdot \\
		 & \quad	\begin{pmatrix}
		               e^{i\omega}   &  0 \\
							     0             &  e^{-i\omega}
		        \end{pmatrix}
						\begin{pmatrix}
	               \sec\theta_{N-1}   &  -\tan\theta_{N-1} \\
						     -\tan\theta_{N-1}  &  \sec\theta_{N-1}
		        \end{pmatrix}
						\cdots \\
			& \quad		\begin{pmatrix}
		               \sec\theta_{1}   &  -\tan\theta_{1} \\
							     -\tan\theta_{1}  &  \sec\theta_{1}
		        \end{pmatrix}
						\begin{pmatrix}
		               e^{i\frac{\omega}{2}}   &  0 \\
							     0                       &  e^{-i\frac{\omega}{2}}
		        \end{pmatrix}
						\tmatrix{e^{i\frac{\omega}{2}}}{e^{-i\frac{\omega}{2}}}.
	\end{split}
\end{equation}
Replace $\tmatrix{e^{-i\frac{\omega}{2}}}{e^{i\frac{\omega}{2}}}$ and
$\tmatrix{e^{i\frac{\omega}{2}}}{e^{-i\frac{\omega}{2}}}$ in
Eq.~\eqref{RecurRelaBN} with the identities
\begin{equation}
   \begin{split}
     \tmatrix{e^{-i\frac{\omega}{2}}}{e^{i\frac{\omega}{2}}}
		  & \equiv \begin{pmatrix}
		               e^{-i\frac{\omega}{2}}   &  0 \\
							     0                       &  e^{i\frac{\omega}{2}}
		        \end{pmatrix} \tmatrix{1}{1}, \\
		 \tmatrix{e^{i\frac{\omega}{2}}}{e^{-i\frac{\omega}{2}}}
		  & \equiv \begin{pmatrix}
		               e^{i\frac{\omega}{2}}   &  0 \\
							     0                       &  e^{-i\frac{\omega}{2}}
		        \end{pmatrix} \tmatrix{1}{1},
	 \end{split}
\end{equation}
Eq.~\eqref{RecurRelaBN} then becomes
\begin{equation}
     \label{RecurRelaBN2}
		\begin{split}
		 \tmatrix{1}{1}
      & = c
		        \begin{pmatrix}
		               e^{i\omega}   &  0 \\
							     0                       &  e^{-i\omega}
		        \end{pmatrix}	\cdot \\
			& \quad
						\prod^{N}_{n=1}{\left[ \begin{pmatrix}
		               \sec\theta_n   &  -\tan\theta_n \\
							     -\tan\theta_n  &  \sec\theta_n
		        \end{pmatrix}
						\begin{pmatrix}
		               e^{i\omega}   &  0 \\
							     0                       &  e^{-i\omega}
		        \end{pmatrix}\right]}
						\tmatrix{1}{1}.
		\end{split}
\end{equation}
Multiply matrix $P^{-1}$ from the left of both sides of Eq.~\eqref{RecurRelaBN2}
and insert the identity $I=P^{-1}P$ between neighboring matrices in the right hand side, where
$P^{-1}=P=(\sigma_x + \sigma_z)/\sqrt{2}$,
we will arrive at Eq.~\eqref{RecurRelaBN3} because
\begin{equation}
   \begin{split}
      P^{-1} \begin{pmatrix}
		               e^{i\omega}   &  0 \\
							     0                       &  e^{-i\omega}
		        \end{pmatrix}	
		  P & =   \begin{pmatrix}
		               \cos \omega   &  i\sin \omega \\
							    i\sin \omega   &  \cos \omega
		        \end{pmatrix},	 \\
			P^{-1} \begin{pmatrix}
		               \sec\theta_n   &  -\tan\theta_n \\
							     -\tan\theta_n  &  \sec\theta_n
		        \end{pmatrix}
		  P & =   \begin{pmatrix}
		               \tan \vt_n   &  0 \\
							     0            &  \tan \vt_n
		        \end{pmatrix},
	\end{split}
\end{equation}
where $\vt_n = \frac{\pi}{4} - \frac{\theta_n}{2}$.

\section{More on the boundary conditions}
\label{sec:MB}
Previously we employ one specific boundary condition to study
the physics of ODD,
but leave three other boundary conditions unexplored. Here we will
briefly summarize the special quasi-energies and the corresponding
states~\cite{Kitagawa2012,Asboth2012} for these different boundary conditions.
Given the bulk $\theta_n=\pi/4+\delta_n$ with $|\delta_n|<\pi/4$,
then the boundary condition $(\theta_0,\theta_{N+1})=(-\pi/2,\pi/2)$ [$(\pi/2,-\pi/2)$]
will lead to the edge states with quasi-energy $\omega=0\text{ or }\pi$ localized
around the boundary site $n=0$ [$n=N+1$]. For convenience, we assume $\delta_n=0$ in our
qualitative discussions below.

Interestingly, the $0\text{ or }\pi$ quasi-energy states are absent under
the boundary conditions $(\theta_0,\theta_{N+1})=(\pi/2,\pi/2)$.
For the case of $(\theta_0,\theta_{N+1})=(-\pi/2,-\pi/2)$,
it can be shown that there exist localized edge states with quasi-energies slightly differing from
$0$ or $\pi$. These features are also relevant to understand the topological properties in
QW~\cite{Kitagawa2012,Asboth2012}. Here we elaborate these features using the transfer matrix formalism (TMF).
Following the same method in Sec.~\ref{sec:EDOS}, the relation between
2 boundaries given by Eq.~\eqref{RecurRela}
can be written in the form analogous to Eq.~\eqref{RecurRelaBN3}:
\begin{equation}
       \label{RecurRelaBND}
		 \tmatrix{1}{0}
      = c_a
		        \begin{pmatrix}
		               \cos \omega   &  i\sin \omega \\
							    i\sin \omega   &  \cos \omega
		        \end{pmatrix}					
						\cdot \bP \cdot
						\tmatrix{0}{i},
\end{equation}
\begin{equation}
       \label{RecurRelaBND2}
		 \tmatrix{0}{i}
      = c_b
		        \begin{pmatrix}
		               \cos \omega   &  i\sin \omega \\
							    i\sin \omega   &  \cos \omega
		        \end{pmatrix}					
						\cdot \bP \cdot
						\tmatrix{1}{0}.
\end{equation}
Here $\vt_n=\frac{\pi}{4} - \frac{\theta_n}{2}$ and $\bP$ is given
in Eq.~\eqref{RecurRelaBN3}. Eq.~\eqref{RecurRelaBND} is for the
boundary condition $(\theta_0,\theta_{N+1})=(\pi/2,\pi/2)$ and Eq.~\eqref{RecurRelaBND2}
is for $(\theta_0,\theta_{N+1})=(-\pi/2,-\pi/2)$.

In the case of Eq.~\eqref{RecurRelaBND} and using the same language as in Sec.~III A,
an actual quasi-energy $\omega$ needs to bring a vector initially
at the $y$-axis, $\tmatrix{0}{i}$ to the $x$-axis, $\tmatrix{1}{0}$.
For simplicity, we assume the vector goes from the positive $y$-axis to
the negative $x$-axis.
$\omega=0$ or $\pi$ certainly cannot accomplish this task since it will
let the vector stay in $y$-axis. Let us check if a small value $\epsilon$ which
slightly above 0 can be the quasi-energy,   using Eq.~\eqref{RelaTan} with $\theta_n=\pi/4$,
$\phi_1 = \pi/2$, $\phi_N=\pi-\epsilon$ and $\vt_n = \pi/8$. It then follows
that $\tan\phi_n$ should approach 0 from $-\infty$ (that is, after the vector enters the second quadrant).
However,  this cannot be true since $\cot^2(\pi/8)>>1$ will prevent $\tan\phi_n$
from approaching 0.  
Together with other simple considerations, it is seen that under the above boundary condition,
$\omega=0$, $\pi$ and any value near them cannot be the quasi-energies of the system.

In the case of Eq.~\eqref{RecurRelaBND2}, the vector
should go from the $x$-axis to the $y$-axis. For simplicity,
we assume the vector goes from the positive $x$-axis to
the positive $y$-axis. This corresponds to
$\tan\phi_n$ going from 0 to $\infty$. It is obvious that $\omega=0$ or $\pi$
cannot achieve this goal. Again we consider a small value $\omega=\epsilon$.
Now the factor $\cot^2(\pi/8)>>1$ in Eq.~\eqref{RelaTan} will speed up this process,
thus indicating that a small $\omega=\epsilon$  may satisfy Eq.~\eqref{RecurRelaBND2}.
In addition,  according to
Fig.~\ref{fig:QW_VectorsInterpretation}, when $\phi$ is smaller than $\pi/4$,
the length of the vector tends to decrease exponentially, and after it passes $\pi/4$,
the length starts to increase exponentially. Therefore, the corresponding
eigenstate is sharply localized at both edges.
Except for this particular $\epsilon$, we may expect
that a vector with a slightly larger $\omega$ may pass two
more quadrants to reach the negative $y$-axis such that it can be
another quasi-energy of the system. But
this is not true because the vector cannot goes from
the positive $y$-axis to the negative $x$-axis.
Hence, this small quasi-energy $\epsilon$ is well-separated from other quasi-energies.
Until a quasi-energy $\omega$ becomes large enough to cross the 2nd quadrant (i.e.,
from the positive $y$-axis to the negative $x$-axis),
no other $\omega$ can satisfy Eq.~\eqref{RecurRelaBND2}.


\section{Other special quasi-energies in the disordered QW}
\label{sec:QW_other_quasi}
Obuse {\it et al} ~\cite{Obuse2011}  numerically showed that $\omega=\pm\pi/2$ can be also
special quasi-energy values with singular DOS, which hence indicate the presence of ODD
in disordered QW.
Here we use the method developed in Sec.~\ref{sec:EDOS} to discuss these special quasi-energy values.x

We start with Eqs.~\eqref{TMM_U} and \eqref{DTQW_spinor} in Sec.~\ref{sec:QW_solving}.
Without loss of generality, we choose $\omega=\pi/2$.  
Then the chain relation
analogous to Eq.~\eqref{RecurRela} will be
\begin{equation}
     \label{RecurRela_other}
		\begin{split}
       c_{N} \tmatrix{-\sin\theta_0}{i} & = \prod^{N}_{n=1}{T_{n}} \cdot c_0 \tmatrix{i}{\sin\theta_{N+1}} \text{ with  } \\
			T_n & = i\sigma_z\sec\theta_n - \sigma_x \tan\theta_n.
		\end{split}
\end{equation}
Define
\begin{equation}
   \label{QW_pair}
   P_m\equiv T_{2m}\cdot T_{2m-1},
\end{equation}
so
\begin{equation}
   \begin{split}
     P_m & = \left(\tan\theta_{2m} \tan\theta_{2m-1} -\sec\theta_{2m}\sec\theta_{2m-1}\right) \cdot \bI +\\
		& \quad \left(\sec\theta_{2m} \tan\theta_{2m-1} -\tan\theta_{2m}\sec\theta_{2m-1}\right) \cdot \sigma_y.
		\end{split}
\end{equation}
Expressing $P_m$ in the basis of $\sigma_y$, we have
\begin{equation}
     P_m = \begin{pmatrix}
		               -\cot\vt_{2m} \tan\vt_{2m-1}  &  0 \\
							     0 &  -\tan\vt_{2m} \cot\vt_{2m-1}
		    \end{pmatrix},
\end{equation}
where $\vt_j = \frac{\pi}{4} - \frac{\theta_j}{2}$.
So in the $\sigma_y$ basis for even $N$,
\begin{equation}
      \prod^{N}_{n=1}{T_{n}} = \begin{pmatrix}
		               \lambda_+  &  0 \\
							     0 &  \lambda_-
		    \end{pmatrix}
\end{equation}
with
\begin{equation}
   \label{QW_lambda}
     \lambda_+=\lambda^{-1}_-=(-1)^{\frac{N}{2}} \cot\vt_{N}
		  \tan\vt_{N-1} \cdots \cot\vt_{2}  \tan\vt_{1}.
\end{equation}
Returning to the $\sigma_z$ basis, we have
\begin{equation}
   \label{QW_TMM_prod}
   \prod^{N}_{n=1}{T_{n}}=\frac{1}{2} \left[(\lambda_+ + \lambda_-)\cdot \bI + (\lambda_+ - \lambda_-)\cdot \sigma_y\right].
\end{equation}
We substitute Eq.~\eqref{QW_TMM_prod} into Eq.~\eqref{RecurRela_other}
and find that the boundary conditions $\theta_0 = \theta_{N+1} = \pm \pi/2$
will make Eq.~\eqref{RecurRela_other} hold, while $\theta_0=\pi/2$, $\theta_{N+1}=-\pi/2$
or $\theta_0=-\pi/2$, $\theta_{N+1}=\pi/2$ cannot. This conclusion is independent
of the actual values of $\theta_n$ ($n=1,2\cdots N$), so whether $\omega=\pi/2$
is the quasi-energy of the system is determined by the boundary conditions,
as well as the parity of the number of system sites.

In our set-up, $N+2$ is the total number of sites in the
disordered QW chain (See Fig.~\ref{fig:QW_setup}).
Each bulk site corresponds to one transfer matrix, and totally
$N$ transfer matrices are involved in the calculation.
When $N$ is odd, one transfer matrix will be left if we pair
those transfer matrices according to Eq.~\eqref{QW_pair}. This leads to
\begin{equation}
   \label{QW_TMM_prodII}
	 \begin{split}
      \prod^{N}_{n=1}{T_{n}}&=\frac{1}{2} \left(i\sigma_z\sec\theta_N - \sigma_x \tan\theta_N\right)\cdot \\
			& \quad \left[(\lambda'_+ + \lambda'_-)\cdot \bI + (\lambda'_+ - \lambda'_-)\cdot \sigma_y\right],
	 \end{split}
\end{equation}
where $\lambda'_+$ and $\lambda'_-$ are obtained from Eq.~\eqref{QW_lambda} by
substituting $N$ with $N-1$. Different from the case of even $N$, the additional $\sigma_x$
and $\sigma_z$ flip the eigen spinors of $\sigma_y$, resulting in the opposite conclusions. In particular,
boundary conditions $\theta_0=\pi/2$, $\theta_{N+1}=-\pi/2$ or $\theta_0=-\pi/2$, $\theta_{N+1}=\pi/2$
will give rise to $\omega=\pi/2$, while $\theta_0 = \theta_{N+1} = \pm \pi/2$ cannot.

We summarize the results in the Tab.~\ref{tab:BC_II}. Those states with exactly
quasi-energy $\pm\pi/2$ are delocalized. For example, in the case of even $N$
and $\theta_0=\theta_{N+1}=-\pi/2$, we substitute Eq.~\eqref{QW_TMM_prod} into
Eq.~\eqref{RecurRela_other} and get
\begin{equation}
      c_{N} \tmatrix{1}{i} = i c_0 \lambda_+\tmatrix{1}{i}.
\end{equation}
Therefore, the spinors at two boundaries are the eigen spinor
of $\sigma_y$, and they are connected by $\lambda_+$ in
Eq.~\eqref{QW_lambda}. 
In general $\lambda_+\approx 1$ because $\cot\vt_{j}$
and $\tan\vt_{k}$ ($j,k\in[1,N]$ are arbitrary indices)
will approximately cancel each other given that $\theta_{j\backslash k}$
are drawn randomly from a given distribution. This
resembles the 0-mode in Sec.~\ref{sec:zeroE}.
Note that, the delocalized 0-mode requires $\theta_n$ to be drawn from
a distribution symmetric with respect to $\theta=0$ (we choose $\theta_n\in[-\tw,\tw]$ in our study),
whereas the delocalized $\pm\pi/2$ states do not have this constraint.
However, the advantage of a delocalized state at $\omega=0$ is that it can be obtained from localized
$\omega=0$ state through an adiabatic protocol (See Sec.~\ref{sec:QW_expt}).
By contrast, the $\omega=\pm\pi/2$ states cannot be obtained in this manner.  The reason is simple.
States with $\omega=\pm\pi/2$ are delocalized
regardless of $\avg{\theta}$, the mean value of $\theta_n$; 
whereas a delocalized $\omega=0$ state requires $\avg{\theta}\approx 0$.
\begin{table}
\medskip
\begin{tabular}{|c|c|c|}
\hline
  Boundary condition & $\omega=\pm\frac{\pi}{2}$, $N$ even & $\omega=\pm\frac{\pi}{2}$, $N$ odd \\ \hline
  $\theta_0=\frac{\pi}{2}=\theta_{N+1}$ & Y & N \\ \hline
  $\theta_0=-\frac{\pi}{2}$, $\theta_{N+1}=\frac{\pi}{2}$  & N & Y \\ \hline
	$\theta_0 = \frac{\pi}{2}$, $\theta_{N+1} = -\frac{\pi}{2}$ & N & Y \\ \hline
	$\theta_0=-\frac{\pi}{2}=\theta_{N+1}$ & Y & N\\ \hline
\end{tabular}
\caption{The existence (Y)  or nonexistence (N) of $\pm \frac{\pi}{2}$ modes under
 different boundary conditions. In the bulk, values of $\theta_n$ ($1\leq n\leq N$)
are assumed not to satisfy $\pi/4-\theta_n/2 = j\cdot \pi/2$ ($j$ is an integer).
		}
\label{tab:BC_II}
\end{table}


%

\end{document}